%% file: proceedings.tex
\documentclass{sigchi}

\toappear{\scriptsize Permission to make digital or hard copies of all or part of this work for personal or classroom use is granted without fee provided that copies are not made or distributed for profit or commercial advantage and that copies bear this notice and the full citation on the first page. Copyrights for components of this work owned by others than ACM must be honored. Abstracting with credit is permitted. To copy otherwise, or republish, to post on servers or to redistribute to lists, requires prior specific permission and/or a fee. Request permissions from permissions@acm.org. \\
{\emph{CHI '20, April 25--30, 2020, Honolulu, HI, USA.} } \\
\copyright~2020 Association for Computing Machinery. \\
ACM ISBN 978-1-4503-6708-0/20/04\ ...\$15.00. \\
http://dx.doi.org/10.1145/3313831.3376610 }
\usepackage{balance}       
\usepackage{graphics}      
\usepackage[T1]{fontenc}   
\usepackage{txfonts}
\usepackage{mathptmx}
\usepackage[pdflang={en-US},pdftex]{hyperref}
\usepackage{color}
\usepackage{booktabs}
\usepackage{textcomp}
\usepackage{breqn}

\usepackage{microtype}        
\usepackage{ccicons}          

\usepackage{todonotes}
\usepackage{CJKutf8}

\usetikzlibrary{calc}
\def\BibTeX{{\rm B\kern-.05em{\sc i\kern-.025em b}\kern-.08emT\kern-.1667em\lower.7ex\hbox{E}\kern-.125emX}}

\def\plaintitle{ORCSolver: An Efficient Solver for Adaptive GUI Layout with OR-Constraints}
\def\plainauthor{Yue Jiang, Wolfgang Stuerzlinger, Matthias Zwicker, Christof Lutteroth}
\def\plainkeywords{GUI builder, layout manager, constraint-based layout, visual interface design, visual programming, optimization}

\makeatletter
\def\url@leostyle{%
  \@ifundefined{selectfont}{
    \def\UrlFont{\sf}
  }{
    \def\UrlFont{\small\bf\ttfamily}
  }}
\makeatother
\def\pprw{8.5in}
\def\pprh{11in}

\usepackage[ruled,vlined,linesnumbered]{algorithm2e}

\definecolor{linkColor}{RGB}{6,125,233}
\hypersetup{%
  pdftitle={\plaintitle},
  pdfauthor={\plainauthor},
  pdfkeywords={\plainkeywords},
  pdfdisplaydoctitle=true, 
  bookmarksnumbered,
  pdfstartview={FitH},
  colorlinks,
  citecolor=black,
  filecolor=black,
  linkcolor=black,
  urlcolor=linkColor,
  breaklinks=true,
  hypertexnames=false
}

\begin{document}

\title{\plaintitle}

\numberofauthors{1}
\author{%
\alignauthor{Yue Jiang$^1$~~~~Wolfgang Stuerzlinger$^2$~~~~Matthias Zwicker$^1$~~~~Christof Lutteroth$^3$\\
    \affaddr{$^1$Department of Computer Science, University of Maryland, College Park, MD, USA}\\
    \affaddr{$^2$School of Interactive Arts + Technology (SIAT), Simon Fraser University, Vancouver, BC, Canada}\\
    \affaddr{$^3$Department of Computer Science, University of Bath, Bath, UK}\\
    \email{\{yuejiang, zwicker\}@cs.umd.edu~~~~w.s@sfu.ca~~~~c.lutteroth@bath.ac.uk}}
}

\teaser{
\centering
  \includegraphics[width=0.8\textwidth]{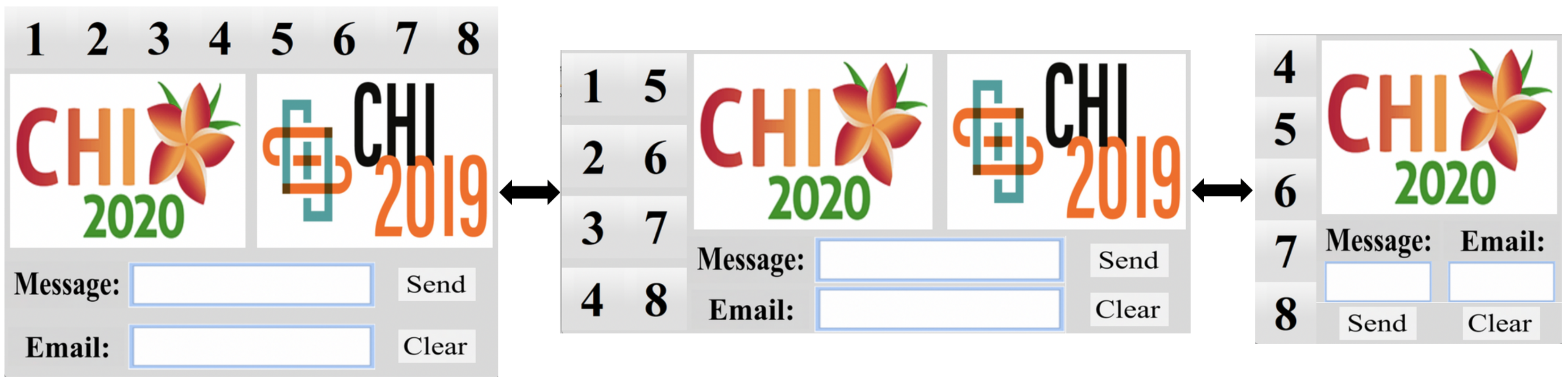}
\caption{ORCSolver is able to adapt a layout between different sizes and orientations at {near-interactive} rates, based on a single layout specification. The layout is adjusted to fit the aspect ratio, and the optional ``CHI2019'' logo and buttons 1, 2, 3 are automatically removed as space gets scarce.}
\label{fig:teaser}
}

\maketitle

\begin{abstract}
OR-constrained (ORC) graphical user interface layouts unify conventional constraint-based layouts with flow layouts, which enables the definition of flexible layouts that adapt to screens with different sizes, orientations, or aspect ratios with only a single layout specification. Unfortunately, solving ORC layouts with current solvers is time-consuming and the needed time increases exponentially with the number of widgets and constraints. To address this challenge, we propose ORCSolver, a novel solving technique for adaptive ORC layouts, based on a branch-and-bound approach with heuristic preprocessing. We demonstrate that ORCSolver simplifies ORC specifications at runtime and our approach can solve ORC layout specifications efficiently at near-interactive rates.
\end{abstract}

\ccsdesc[500]{Human-centered computing~User interface toolkits}

\keywords{\plainkeywords}

\printccsdesc

\input{00-introduction}
\input{01-related_work}
\input{02-formalization}
\input{03-overview}
\input{04-algorithm}
\input{05-implementation}

\input{06-evaluation}

\input{07-conclusion}

\balance{}



\bibliographystyle{SIGCHI-Reference-Format}
\bibliography{proceedings}

\end{document}

%% file: 00-introduction.tex
\section{Introduction}

Constraint-based layout models have been widely used in many graphical user interfaces (GUIs) for applications because they are more flexible and powerful than other layout models. For instance, Apple's AutoLayout \cite{sadun2013ios} adapts interfaces on different devices ranging from desktops to smartphones and CSS's Flex(ible) Box\footnote{CSS Flex Box: \url{https://w3schools.com/css/css3_flexbox.asp}} uses constraints to fit the content and solve alignment problems dynamically.
Constraint-based layout models can align widgets across different groups, which is impossible with grid layouts. 
Yet, common constraint-based layout models are not without limitations. They cannot support flow layouts, and despite their flexibility, device diversity has been a long-term challenge for them. Even with constraint-based layouts, separate layout specifications need to be defined for different sizes, orientations (portrait and landscape), or aspect ratios of the screens. Such multiple specifications can be hard to maintain and to keep consistent. 

To address these problems, Jiang et al.\ \cite{jiang2019ORC} proposed an approach for constraint-based graphical user interface layouts based on OR-constraints (ORC). An OR-constraint is a disjunction of multiple constraints, where only one needs to be true. ORC layouts specify adaptive GUIs for different devices, as they can adapt a GUI to screens with different sizes, orientations, and aspect ratios with predictable results, using only a single layout specification. This has the potential to make GUI design more efficient and less error-prone, as designers do not need to manually synchronise changes between different layout specifications for different devices.
However, solving ORC layouts is currently slow, due to the computational complexity involved. Solving OR- and soft constraints is more challenging than solving hard constraints due to the combinatorial explosion. Modern constraint solvers such as Z3 \cite{de2008z3} cannot handle OR-constraints efficiently, as they need to evaluate very large numbers of potential combinations. Although Jiang et al.\ \cite{jiang2019ORC} stated that ORC specifications can be solved interactively, the solving time in their experiments increased dramatically with the number of widgets. For example, their approach took almost 2 seconds to solve a layout with only 30 widgets. 

Here, we propose ORCSolver -- a novel efficient solving technique for adaptive GUI layout with OR-constraints. 
ORCSolver is the first solver that can solve complex, realistic layouts made of linear constraints, flows, and their combinations at near-interactive rates.
ORCSolver uses a heuristic optimization algorithm to reduce the computational complexity of solving adaptive GUI layouts with OR-constraints. Instead of feeding many OR- and soft constraints directly into a general-purpose solver, such as Z3 \cite{de2008z3}, we use a branch-and-bound algorithm with efficient heuristics to reduce the solution space and discard branches based on bounding criteria. We then produce a smaller set of constraints, which represent the same layout as the original set but are tailored to the current screen size and have thus drastically reduced complexity. We demonstrate that our solving technique can significantly reduce the solving time for adaptive GUI layouts with OR-constraints, making the following contributions:

%


\begin{itemize}

\item A formal notation for ORC layouts, which can be fed into our ORCSolver and help to formalise solving strategies.

\item A novel solving technique for ORC layouts, based on a branch-and-bound approach with interval arithmetic and modular heuristics, which drastically reduces solving time.

\item A set of heuristics to reduce the number of constraints in ORC layout specifications during solving. We formalize the solver as a framework with pluggable heuristics.

\item Experimental evidence of the efficiency of our approach through comparison with three other solvers, which demonstrates that ORC layouts can be solved much more efficiently ({\em e.g.}, more than two orders of magnitude faster than Z3 with more than 100 widgets).

\end{itemize}

%% file: 01-related_work.tex
\section{Background}

Linear constraint layouts have been widely used for responsive web design \cite{harb2011responsive, marcotte2017responsive} and mobile apps \cite{sahami2013insights, zanden1990automatic}. Constraint-based layouts are easier to maintain compared to programming approaches \cite{myers2000past, Myers1995user} and overcome common limitations of other
layout models, such as an inability to align widgets across containment hierarchies \cite{lutteroth2008modular} and maintenance issues \cite{lutteroth2006user, zeidler2012comparing}. 
Constraint-based layout specifications are usually systems of linear equations and inequalities, being either hard or soft constraints.

{\bf \emph{Soft and Hard Linear Constraints:}}
Hard constraints are ones which must be satisfied, while soft constraints can be neglected if not all constraints can be satisfied simultaneously. Soft constraints form a hierarchy, where each has a weight to define its priority \cite{borning1992constraint}. Higher weights are given to more important widgets, which implies higher priority. Hard constraints are soft constraints with infinite weights.

{\bf \emph{OR-Constraints:}}
OR-constraints are disjunctive, where the entire constraint functions as a hard one, while each disjunctive part is a soft constraint. Only one disjunctive part needs to be satisfied to make the OR-constraint true. OR-constraints are more challenging to solve than hard constraints. For OR-constraints, solvers typically need to check multiple cases to identify a branch in each disjunctive constraint that can be satisfied, while for soft constraints, their attached weights necessitate decisions about which ones should be neglected and which ones satisfied.

{\bf \emph{Interval Arithmetic:}}
Interval arithmetic is useful for solving systems of equations and inequalities \cite{neumaier1991}.
This numerical method puts bounds on each variable to identify ranges that contain promising solutions, to quickly identify valid solutions for complex optimization problems.
When we apply a function $f$ to an uncertain value $x$, the result is uncertain since the input $x$ is indeterminate. Thus, we do not know what the result is or how far we are from it. Instead of using $x$ as the input, we work with an interval $[a,b]$ that contains $x$. Since all its operations are closed, after we apply the function $f$ to the interval $[a,b]$ with interval arithmetic, the resulting interval $[c,d]$ must contain the accurate value of $f(x)$.
We use interval arithmetic to rule out impossible or ineffective solutions. Once we know that the best solution must be in an interval $[c,d]$, we can rule out all the potential solutions outside of that interval.

{\bf \emph{Branch \& Bound (B\&B):}}
The B\&B algorithm  \cite{Markowitz1957} is a well-known approach to solving combinatorial optimization problems. This search algorithm explores the branches of a rooted tree. Each branch contains a subset of the solution set.
With this, B\&B recursively divides the search space into smaller subspaces and minimizes an objective function on all of them. Once it finds that a branch cannot contain an optimal solution, it discards that branch and does not explore its subtree further.
Zeidler et al.\ \cite{zeidler2017generation} used B\&B to automatically generate layouts for alternative screen orientations. They investigated only mobile device rotation, but not different screen sizes. Other layout generation approaches such as SUPPLE \cite{gajos2004supple} also used B\&B. All these approaches suffer from combinatorial explosion.

{\bf \emph{Quadratic Programming (QP) Solvers:}}
QP problems \cite{Frank1956qp} have been widely explored. Most QP solvers are based on three approaches: active set methods \cite{Wolfe1960simplex} ({\em e.g.,} qpOASES \cite{Ferreau2014qpOASES}), interior point methods \cite{Nesterov1994interior} ({\em e.g.,} MOSEK \cite{2017MOSEK}, Gurobi \cite{2016Gurobi}, CVXGEN \cite{Mattingley2012CVXGEN}, OOQP \cite{Gertz2001qp}),  and first order methods \cite{Frank1956qp} ({\em e.g.,} SCS \cite{ODonoghue2016SCS}). The alternating approach of multipliers (ADMM) \cite{Gabay1976ADMM}, a first order method, is widely used in practice for solving QP with good convergence behavior. The OSQP solver \cite{osqp} is a state of the art QP solver using an improved ADMM method \cite{Gabay1976ADMM} to avoid strong dependencies on the problem data.

{\bf \emph{Solving Approaches for OR-Constraints:}}
Previous layout solvers \cite{badros2001cassowary, borning1997solving, Freeman1990incremental, hosobe2000scalable, Hosobe2011HiRise2, jaffar1992clpr, Marriott1998tableau, Marriott2002qoca, sannella1994skyblue, sannella1993multi} can only solve either linear constraint-based layouts or flows but not both, {\em i.e.,} cannot solve systems with OR-constraints. Disjunctive constraints for GUIs were first proposed for non-overlap, {\em i.e.}, to ensure widgets never overlap \cite{Marriott2001disjunctive1}. However, the proposed method cannot solve disjoint disjunctions and works only if solutions can transition continuously between disjuncts without passing through unsatisfiable regions. 
The ORC layout approach \cite{jiang2019ORC} was the first to use the Z3 solver to deal with general GUI layouts with various types of OR-constraints. The Z3 solver \cite{de2008z3} is a satisfiability modulo theory (SMT) solver that is able to solve specifications with disjunctive constraints. It has been previously used for solving CSS layout specifications \cite{panchekha2016automated} and layout editing \cite{hottelier2014programming}. Z3 can handle both disjoint and non-disjoint disjunctions, which enables this approach to unify constraint-based and flow layouts. However, both these approaches suffer from runtime issues.

\section{Related Work}


\subsection{GUI Builders}
GUI builders support designers to specify and generate layouts. FormsVBT \cite{avrahami1989two} introduced a textual representation and an interactive editor for interfaces. Bramble \cite{gleicher1993graphics} applied differential constraints to generate graphical editing applications. Gilt \cite{hashimoto1992graphical} used Graphical Tabs and styles to simplify layout specifications. Amulet \cite{myers1997amulet} allowed developers to combine some properties of flow and grid layouts programmatically. OPUS \cite{hudson1990interactive} supported direct manipulation interfaces with a graphical notation. Ibuild \cite{vlissides1991unidraw} allowed users to modify simulations of layouts. Peridot \cite{myers1986creating}, Druid \cite{singh1990druid}, and Lapidary \cite{zanden1991lapidary} created code automatically based on users' demonstration. UI Fa\c{c}ades \cite{Stuerzlinger2006facades} afforded direct manipulation to manually reconfigure and recombine existing interfaces.
Modern GUI builder approaches can create layout constraints robustly based on direct manipulation \cite{scoditti2009new, zeidler2013auckland, weber2010reduction}. Based on intersections of objects, Rockit \cite{karsenty1993inferring} determined graphical constraints in a 2D scene. Most layouts can be specified as constraint systems \cite{zeidler2017tiling}, and designing a good constraint system is important for adaptive GUI layouts: an ambiguous specification with too few constraints (`underspecification') can lead to unpredictable results, while too many (`overspecification') can lead to conflicts, lack of valid solutions, and much increased solving times. 


\subsection{UI Generators}
Some work proposed to improve GUI customization and adaptation \cite{weld2003automatically}. Fogarty et al. \cite{fogarty2003gadget} supported optimization for interface generation. SUPPLE \cite{Gajos2008decision, gajos2004supple} automatically generated user interfaces by minimizing interface operations to meet screen-size constraints. It was also used to generate GUIs for Ubicomp apps \cite{Gajos2005Fast, Gajos04automatically}, improve personalization,
maintain consistency \cite{gajos2005cross}, and generate interfaces for people with physical disabilities \cite{DeRenzi2008opportunities, Gajos2006automatically, Gajos2007automatically, gajos2008improving}. Arnauld \cite{Gajos2005preference} learned and generated the parameters of cost functions in optimization-based systems. Other tools generated layout alternatives through templates \cite{Jacobs2003document} or modifiable suggestions \cite{sinha2015responsive, zanden1990automatic}.




\subsection{Layout Solvers}
Given a constraint system, a layout solver is needed to determine a solution for widget positions and sizes. There are various solvers for constraint-based GUI specifications, using linear or quadratic programming \cite{badros2001cassowary, bill1992bricklayer, borning1997solving, hosobe2000scalable, lutteroth2008domain, zeidler2013evaluating}. They use objective functions to minimize deviations from an optimal solution, which can improve aesthetics \cite{zeidler2012constraint}.

CLP(R) \cite{jaffar1992clpr} includes constraints in logic programming for linear equalities and inequalities. It gradually optimizes the variables, but is not suitable for interactive use.
DeltaBlue \cite{Freeman1990incremental, sannella1993multi} and Skyblue \cite{sannella1994skyblue} are local propagation constraint solvers and can handle constraint hierarchies, but cannot solve simultaneous constraints. 
Cassowary \cite{badros2001cassowary, borning1997solving}, an incremental constraint solver for user interface applications, solves linear equalities and inequalities with an incremental simplex approach. 
QOCA \cite{borning1997solving, Marriott1998tableau, Marriott2002qoca} applies tableau-based methods to solve constraint hierarchies.
HiRise \cite{hosobe2000scalable} and HiRise2 \cite{Hosobe2011HiRise2} use a simplex method with an LU decomposition-based algorithm and employ ordered constraint hierarchies to solve linear equality and inequality constraints.

These solvers use different penalty functions (comparators) to handle soft constraints. For example, Cassowary \cite{borning1997solving, badros2001cassowary} uses weighted-sum-better, QOCA \cite{borning1997solving, Marriott1998tableau, Marriott2002qoca} uses least-squares-better, and HiRise \cite{hosobe2000scalable} and HiRise2 \cite{Hosobe2011HiRise2} use locally-error-better to solve constraint hierarchies. These solvers support linear constraints with priorities and aim to satisfy as many constraints as possible, subject to priorities. 

%% file: 02-formalization.tex
\section{Formal Notation}

\begin{figure}[t]
\begin{center}
\includegraphics[width=0.7\columnwidth]{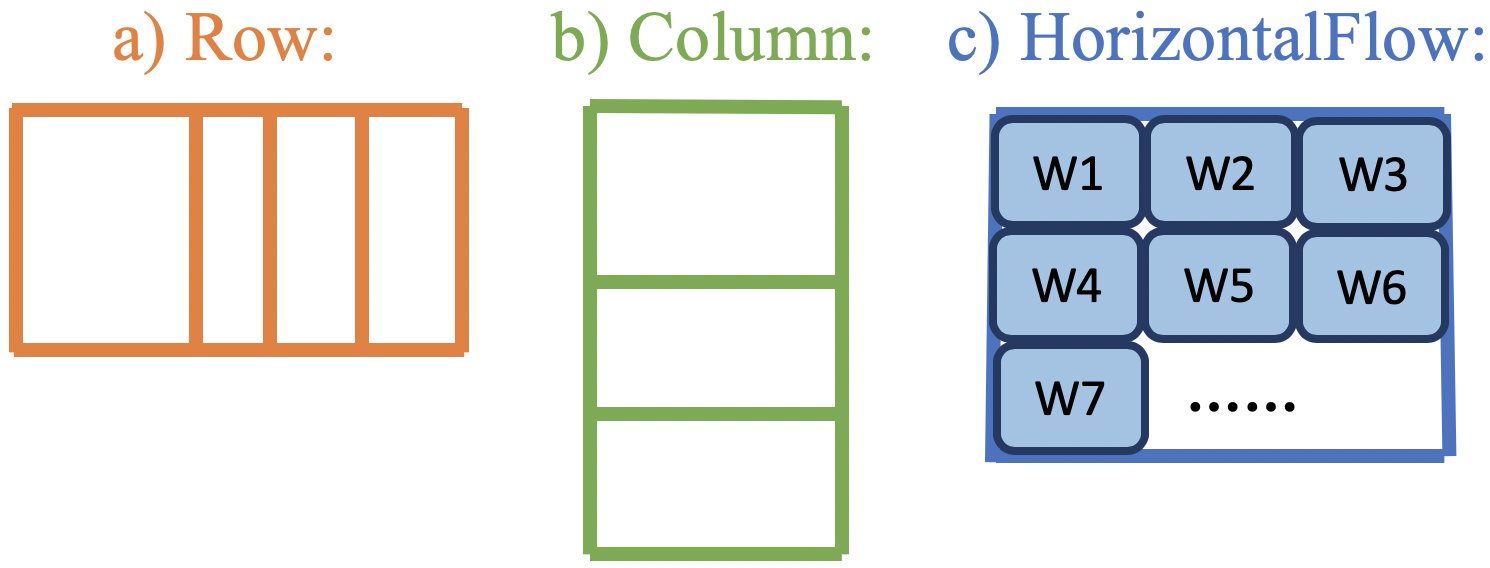}
\end{center}
\caption{a) A Row is a rectangular horizontal arrangement of sub-layout, and b) a Column a rectangular vertical one. c) Horizontal flow.}
\label{fig:api_row_col}
\end{figure}

\begin{figure}[t]
\begin{center}
\includegraphics[width=0.3\columnwidth]{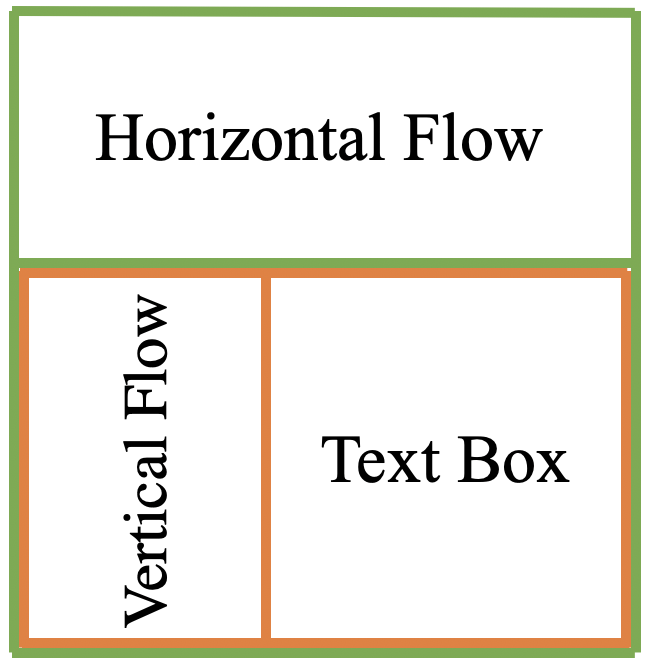}
\end{center}
\caption{ORC layout defined as: {\it Column(HorizontalFlow, Row(VerticalFlow, TextBox))}}

\label{fig:api_ex1}
\end{figure}

\begin{figure}[t]
\begin{center}
\includegraphics[width=0.65\columnwidth]{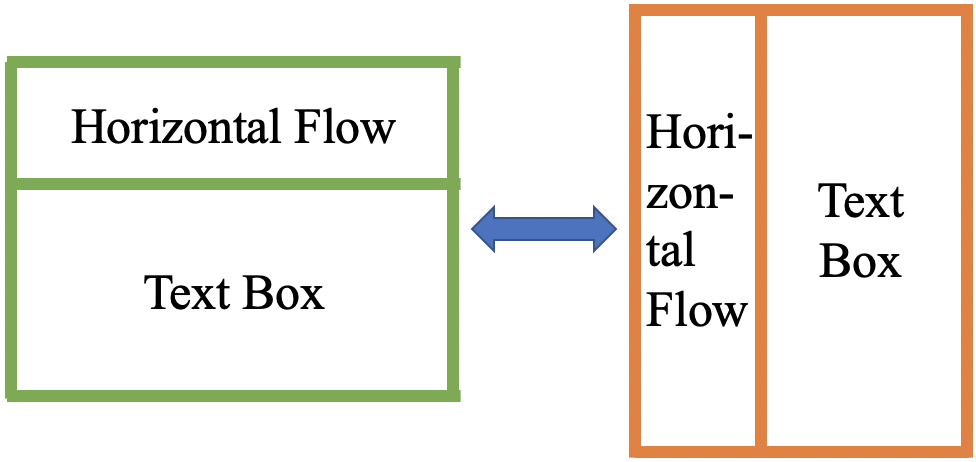}    
\end{center}
\caption{ORC layout defined as: {\it Pivot(Column(HorizontalFlow, TextBox))}}
\label{fig:api_ex2}
\end{figure}

Constraints are low level and hard to specify manually \cite{zeidler2013auckland}; in most cases, layouts can be expressed in a more abstract and developer-friendly manner \cite{zeidler2017tiling}.
We provide a formal notation and corresponding application programming interface (API)\footnote{The full API is included in the open source repository.} to enable GUI developers to specify ORC layouts that can be solved at near-interactive rates by our solver. 
Similar to other GUI toolkits, our notation's layout specifications consist of recursively nested, rectangular layouts that are all subtypes of a generic layout supertype. Layouts can be simple widgets, or full-fledged, modular constraint-based layout specifications \cite{lutteroth2008modular}. Similar to other layout models \cite{Hottelier2014synthesis}, we formalise fixed horizontal and vertical alignments as \textit{Row} and \textit{Column} layout types (\autoref{fig:api_row_col} a, b), and flow layouts as \textit{HorizontalFlow}  (\textit{HF} for short) (\autoref{fig:api_row_col} c) and \textit{VerticalFlow} (\textit{VF} for short). All ORC layout patterns are similarly represented as layout types. For example, the layout in \autoref{fig:api_ex1} can be formalized as \textit{Column(HorizontalFlow, Row(VerticalFlow, TextBox))}. The layout uses a \textit{Column} layout dividing the window into a horizontal flow and the lower part, which is a \textit{Row} layout containing a vertical flow and a text box. 

{\bf \emph{Pivot Layout:}}
{We define a \textit{Pivot} layout to encode more flexibility into the layout design. The \textit{Pivot} layout allows the solver to turn horizontal arrangements into vertical ones, and vice versa, using OR constraints. By default, the \textit{Pivot} layout modifies only its first sub-layout. }
For example, the layout in \autoref{fig:api_ex2} can be formalized as \textit{Pivot(Column(HorizontalFlow, TextBox))}. The {\it HorizontalFlow} could, {\em e.g.}, contain toolbar widgets. 
By using \textit{Pivot}, the {\it Column} can be turned into a {\it Row} to adapt from a landscape orientation (`toolbar at the top') to a portrait orientation (`toolbar on the left'). 
Internally, the solver breaks the layout down into two alternatives: \textit{Column(HorizontalFlow, TextBox)} OR \textit{Row(HorizontalFlow, TextBox)}. 
Figure~\ref{fig:teaser} shows a more complex example with the specification {\it Pivot(Column(HorizontalFlow1, Column(HorizontalFlow2, Pivot(Column(HorizontalFlow3, HorizontalFlow4)))))}, with
the ``CHI2019'' logo and the buttons 1, 2, 3 defined as optional in the layout. 
{The first \textit{Pivot} controls the numbered buttons, and the second \textit{Pivot} independently controls the bottom input fields. If found to be more optimal by the solver, the first \textit{Pivot} could move the numbered buttons to the left while the second \textit{Pivot} does not change the input fields (Figure~\ref{fig:teaser} middle $\rightarrow$ left), or the first \textit{Pivot} could keep the numbered buttons left while the second \textit{Pivot} turns the input fields (Figure~\ref{fig:teaser} middle $\rightarrow$ right).
}

{\bf \emph{Scope:}}
{Our formal notation can represent any rectangular tiling layout, {\em i.e.}, any grid-like subdivision through linear constraints \cite{weber2010reduction, zeidler2017tiling}. Common layouts such as box, grid, and flow are directly represented with our formal notation, and low-level linear constraints can be added where necessary, {\em e.g.,} for alignment across sub-layout boundaries and row-/col-span. In contrast to other layout notations \cite{badros2001cassowary, borning1997solving, Marriott1998tableau, Marriott2002qoca, zeidler2013auckland, zeidler2017tiling} and beyond simple resizing or flows, our notation can also specify how layouts change in response to changing UI sizes.}

%% file: 03-overview.tex
\section{Solving Overview}

{ORCSolver is a conceptually new approach, integrating branch-and-bound, greedy optimization, and quadratic programming, which specifically addresses multi-device UI layouts requiring a particularly high degree of solving flexibility not supported by other solvers, and offers for the first time multi-device layout solving at near-interactive rates. 
ORCSolver is a general contribution as it can be used for any layout that can be represented with linear constraints and/or flows, which includes {almost} all commonly used layout models and document layout patterns \cite{jiang2019ORC, weber2010reduction, zeidler2013auckland}. Some rare layouts are currently not supported, such as overlapping elements \cite{zeidler2017tiling}.} 

Instead of feeding a layout specification with a potentially large number of OR- and soft constraints directly into a constraint solver, ORCSolver uses a B\&B algorithm to keep track of potential solutions for all the sub-layouts and prune branches based on bounding criteria. It also uses modular heuristics for each layout type to reduce the number of OR constraints by adapting sub-layouts to the given layout context, {\it e.g.}, the given available size. The result is a much smaller set of constraints, representing an optimal or near-optimal adaptation of the original constraint system to the given layout context. The resulting system can then be solved much more quickly than the original using an off-the-shelf solver.


\subsection{Branch \& Bound (B\&B)}
{To start, ORCSolver is given a layout specification in our formal notation. The specification can contain high-level layout types as well as low-level constraints.
ORCSolver then searches the layout space with B\&B by considering feasible subsets of constraints for each sub-layout and ruling out whole branches by considering hard constraints and interval bounds of the objective function. To speed up solving, ORCSolver prioritizes branches that are likely to lead to optimal solutions.}


\subsection{Selection of Sub-layout Solving Order}
To reduce assumptions and speed up B\&B, sub-layouts that are more constrained in their size are processed first. This is done by prioritising sub-layouts according to their number of \emph{firm edges}, {\it i.e.}, edges that cannot move (much) in the layout. 
As ORCSolver optimizes a layout, it defines firm edges iteratively. Initially, the layout boundaries are the only firm edges. As ORCSolver processes sub-layouts, their edges become firm as they are placed in the layout with more certainty. Sub-layouts are preferentially resolved or simplified with greedy algorithms if they have a high number of firm edges, which increases certainty and helps narrow down the search space. 
For example, consider a \textit{Row} at the top of an ORC layout: the leftmost sub-layout has at least two firm edges (top and left). Once ORCSolver processes the leftmost sub-layout, new firm edges are generated around that sub-layout and the next sub-layout to the right has now also at least two firm edges. 

\subsection{Greedy Algorithms}
{During B\&B, the aforementioned heuristics are applied to the sub-layouts of an ORC layout. Our heuristics are algorithms that reduce the number of OR-constraints for the different layout types. Optimizing complex sub-layouts, such as flows, with brute-force B\&B search leads to combinatorial explosion. Our heuristics make the B\&B approach computationally tractable, which then also assures an overall optimal layout.}

ORCSolver uses heuristics for the different ORC layout subtypes that reduce the number of OR- and soft constraints by either removing them or changing them into hard constraints, while at the same time ensuring that the simplified constraint system is equivalent (or near-equivalent) to the original one. 
For example, horizontal flow layouts use an OR-constraint ``to the right of the previous widget OR at the beginning of the next row'', which is necessary for generality but introduces one OR-constraint per involved widget.
If ORCSolver can estimate which widgets should be placed into which rows during our constraint reduction phase, it can simply assign either a ``to the right of the previous widget'' or an ``at the beginning of the next row'' constraint to each widget. Both of these are hard constraints and can be solved easily.
Another example is that each widget typically has minimum, preferred, and maximum sizes. Typically, these are implemented as hard constraints that make sure the widget size is larger than or equal to the minimum size and smaller than or equal to the maximum size, and also a soft constraint to preferably give it its preferred size. 
In many cases, ORCSolver can estimate the final size of a widget during our heuristic preprocessing, which means it can then replace all the size constraints of a widget with two simple hard constraints that set its appropriate size.

\subsection{Solver Selection}
When all sub-layouts have been processed with heuristics and OR- and soft constraints have been removed as much as possible, ORCSolver has reached a leaf node of the B\&B search tree. Before backtracking to process sub-layouts with different parameters, 
ORCSolver finishes solving the overall layout by applying a standard solver on the reduced specification. If several OR-constraints could not be eliminated, then ORCSolver uses the Z3 solver as it can efficiently solve general disjunctive constraints. If all OR-constraints could be eliminated, ORCSolver uses a simpler linear solver such as Cassowary \cite{badros2001cassowary} or a quadratic solver such as OSQP \cite{osqp}. If very few OR-constraints remain, ORCSolver traverses all possible alternatives using B\&B and solves them with the simpler solver, as for few alternatives this is often faster than using a more general solver such as Z3.


%% file: 04-algorithm.tex

 \begin{algorithm}[t]
    \SetKwInOut{Input}{Input}
    \SetKwInOut{Output}{Output}
\SetAlgoLined
$i \leftarrow 0$ \tcp {widget index}
\For{$r\leftarrow 1$ \KwTo $numRows$}{
    $start \leftarrow i$ \tcp {index of first widget in row}
    \vspace{1mm}
    
    \tcp {remaining available layout width}
    $width_{totalAvail} \leftarrow (numRows - r) \times width_{row}$
    \vspace{1mm}
    
    \tcp {avg.\ deviation from pref.\ width}
    $\Delta \leftarrow 
    \dfrac{width_{totalAvail} -  \sum_{j\geq start}(width_j)}{numWidgets - start}$
    \vspace{1mm}
    
    \tcp{keep adding widgets while we get closer to row width}
    $width_{rowAvail} \leftarrow width_{row}$
    
    \While{$i < num\_widgets$ AND $|width_{rowAvail} - (prefWidth_i + \Delta)| < |width_{rowAvail}|$} {
        $width_{rowAvail} \leftarrow width_{rowAvail} - (prefWidth_i + \Delta)$
        $i \leftarrow i + 1$
    }
    $end \leftarrow i-1$ \tcp{index of last widget in row}
    \vspace{1mm}
    
    \tcp{avg.\ deviation from pref.\ width for row}
    $\Delta_{row} \leftarrow (width_{row} - \dfrac{\sum_{j=start..end}prefWidth_j}{ end - start + 1}$
    \vspace{1mm}
    
    \tcp{redistribute width in row to fit and set height minimising squared deviations}
    \For{$j\leftarrow start$ \KwTo $end$}{
        $width_j \leftarrow prefWidth_j + \Delta_{row}$
        
        $height_j \leftarrow \dfrac{\sum_{k=start..end}(prefHeight_k)}{end - start + 1}$
    }
  }  
\caption{Greedy Flow Optimisation}
\label{alg:flow}
\end{algorithm}

\begin{table*}[h!]
  \centering
  \begin{tabular}{ | p{0.245\textwidth} | p{0.345\textwidth} | p{0.345\textwidth} |}
    \hline
    \multicolumn{1}{|c|}{Layout Pattern}  & \multicolumn{1}{|c|}{Heuristics} & \multicolumn{1}{|c|}{Time Comparison} \\ \hline

    \begin{minipage}{0.245\textwidth}
    
    \centerline{{\bf Simple Flow}}
    \vspace{1mm}
      \includegraphics[width=45mm]{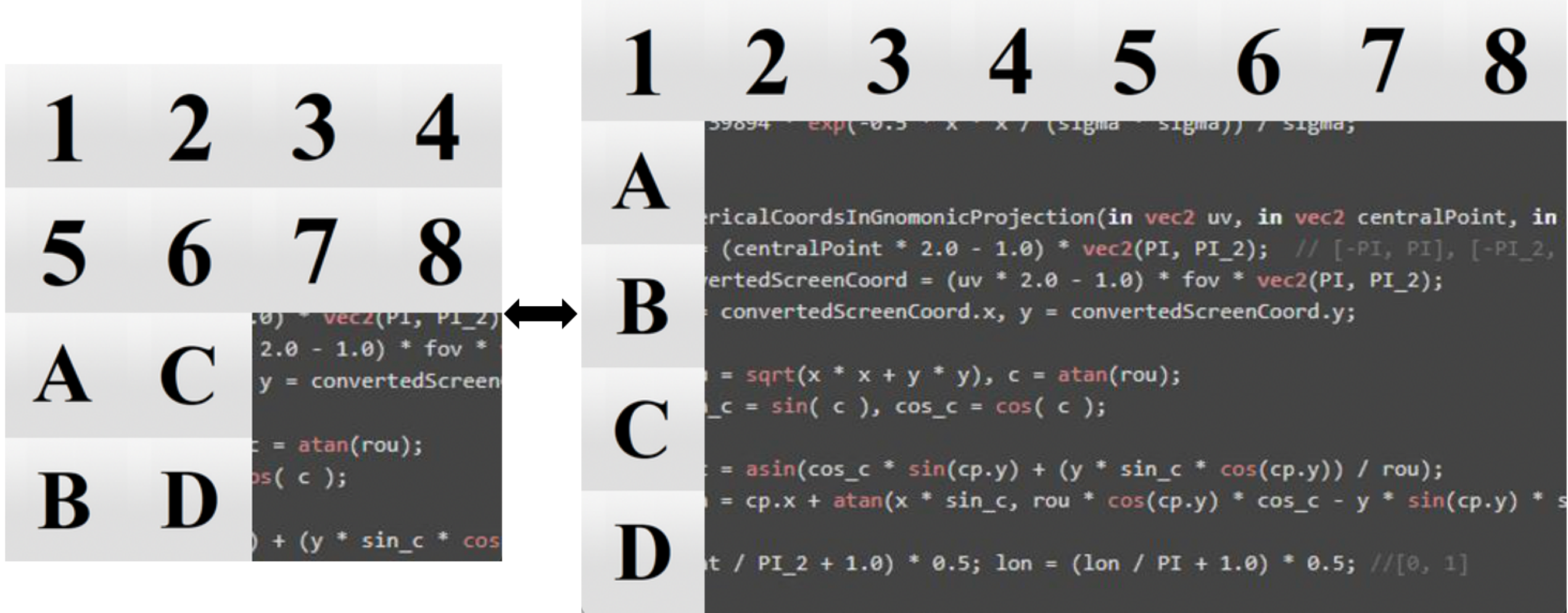}
    \end{minipage}
    &
    \begin{minipage}{0.345\textwidth}
    \vspace{1mm}
   { 
   Given: target number of rows and row widths. See Algorithm 1.
   \begin{enumerate}
   \setlength{\itemsep}{0pt}
  \setlength{\parskip}{0pt}
	\item Calculate the remaining available width in the flow layout (line 4).
	\item Before a new row, compute the average deviation from preferred widths for all remaining widgets in the flow ($\Delta$) to estimate how much width change is needed to fill all rows completely (line 5).
	\item While there are widgets, keep adding them using the next widget's preferred width $prefWidth_i$ plus $\Delta$, which minimises over- \& underfill (lines 6-10).
	\item Calculate average deviation from preferred widths to fill row exactly (line 11).
	\item Adjust widths and heights to minimise the sum of squared deviations from the preferred ones (lines 12-14).
\end{enumerate}
\vspace{1mm}
}

    \end{minipage}
    &
\begin{minipage}{0.345\textwidth}
      \includegraphics[width=63mm]{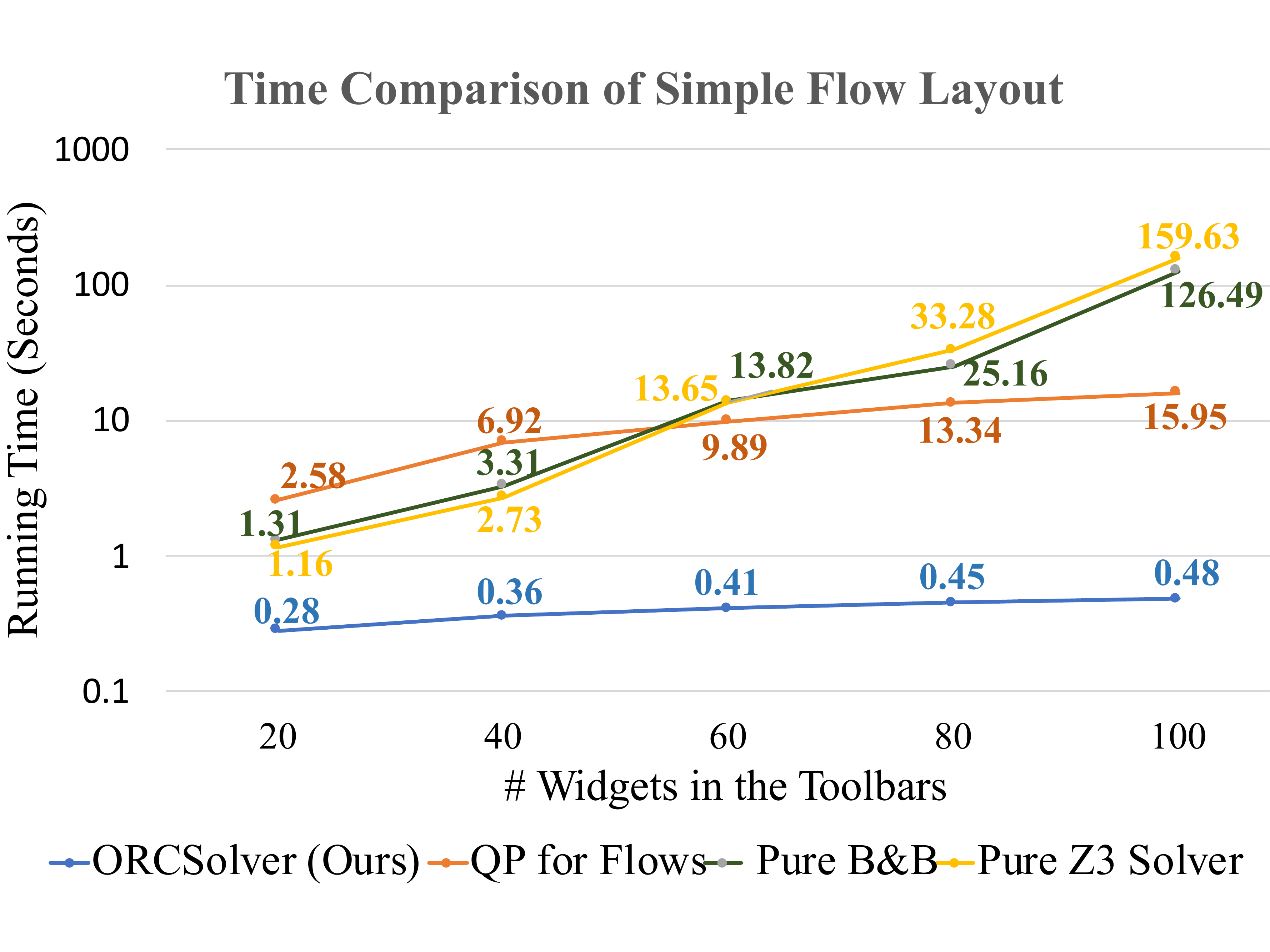}
    \end{minipage}
  
    \\ \hline
    
    \begin{minipage}{0.245\textwidth}
    \centerline{{\bf Connected Flow}}
    \vspace{1mm}
      \includegraphics[width=45mm]{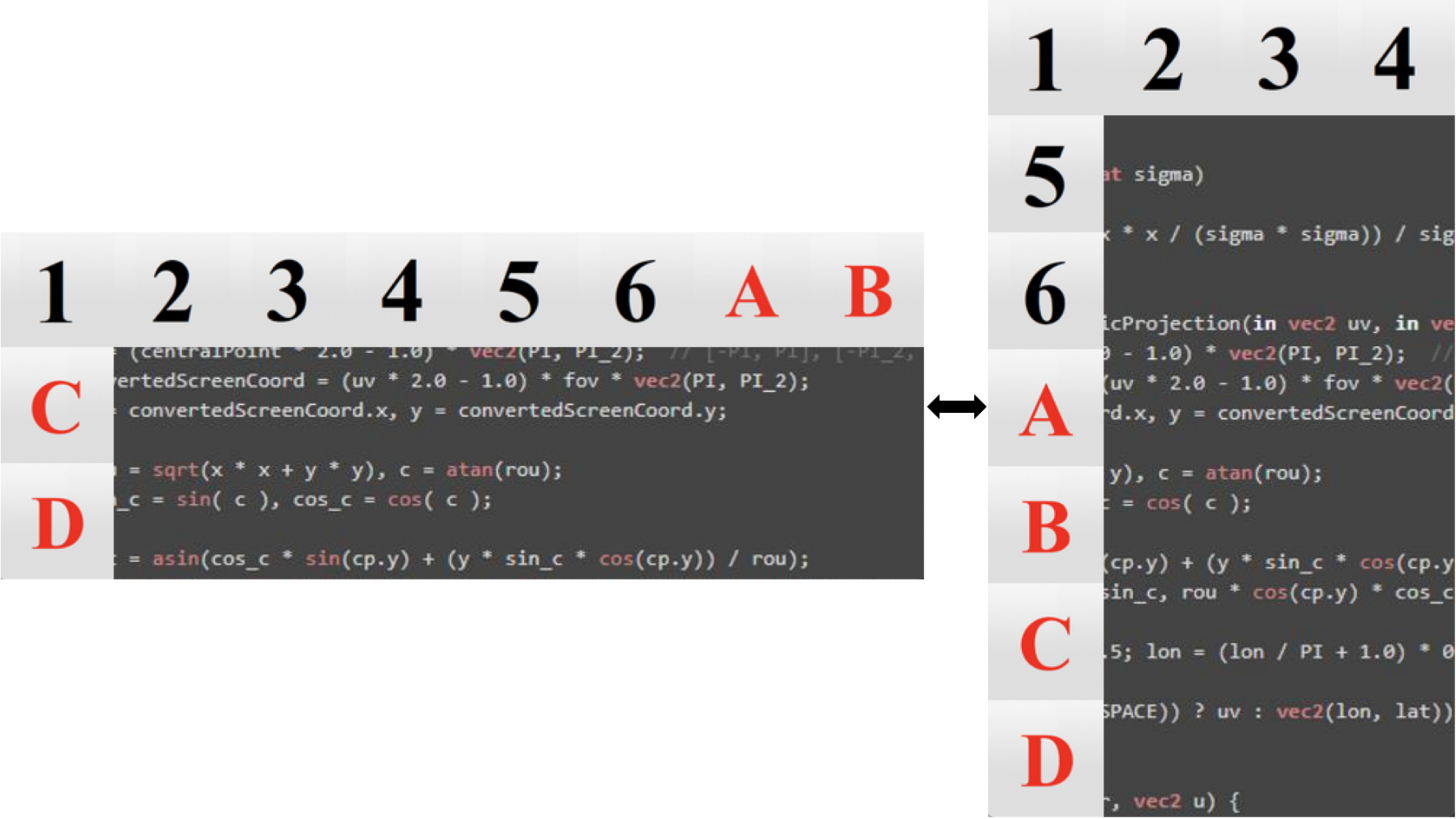}
    \end{minipage}
    
      &  
      \begin{minipage}{0.345\textwidth}
      \vspace{1mm}
      {
      If two flows are connected, widgets in one can move to the other.\\
      Denote the set of widgets in the first flow $W_1$ and second flow $W_2$.\\
   Given: \#rows in the first flow $r \in$ [0, \#rows flowing $W_1 \cup W_2$ in the first flow layout].
   \begin{enumerate}
   \setlength{\itemsep}{0pt}
  \setlength{\parskip}{0pt}
	\item Flow widgets in $W_1\cup W_2$ into the first flow and stop when we reach the end of $r$ rows.
	\item Define final sets of widgets in the 2 flows.\\
	$W_1$' = the widgets put into the first flow.\\
$W_2$' = $(W_1\cup W_2)$ \textbackslash\ $W_1'$.
	\item Apply the flow algorithm to optimize.
\end{enumerate}
\vspace{1mm}
}
    \end{minipage}
    &
\begin{minipage}{0.345\textwidth}
      \includegraphics[width=63mm]{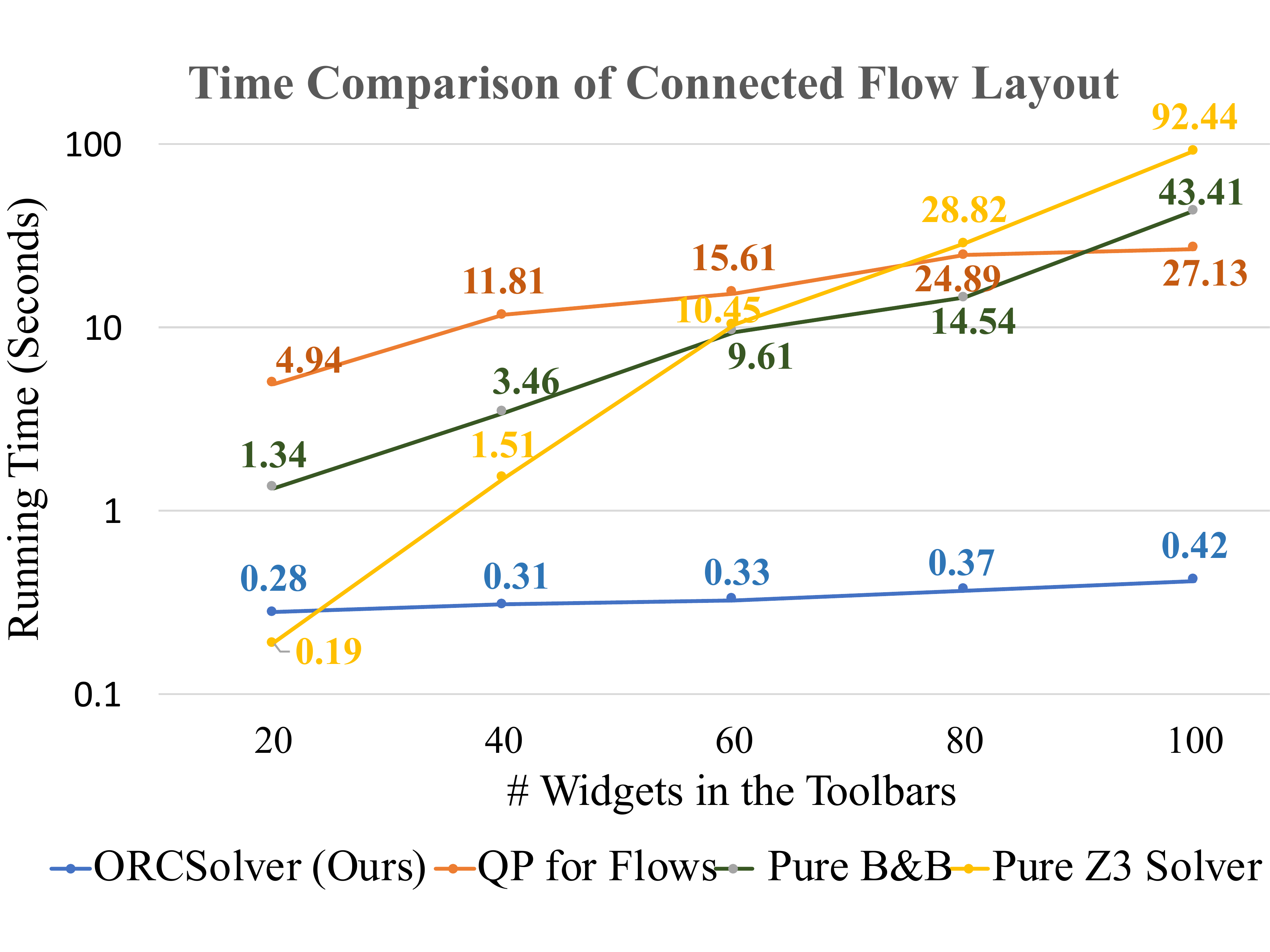}
    \end{minipage}
  
    \\ \hline
    
    \begin{minipage}{0.245\textwidth}
    \centerline{{\bf Flowing Around a Fixed Area}}
    \vspace{1mm}
      \includegraphics[width=45mm]{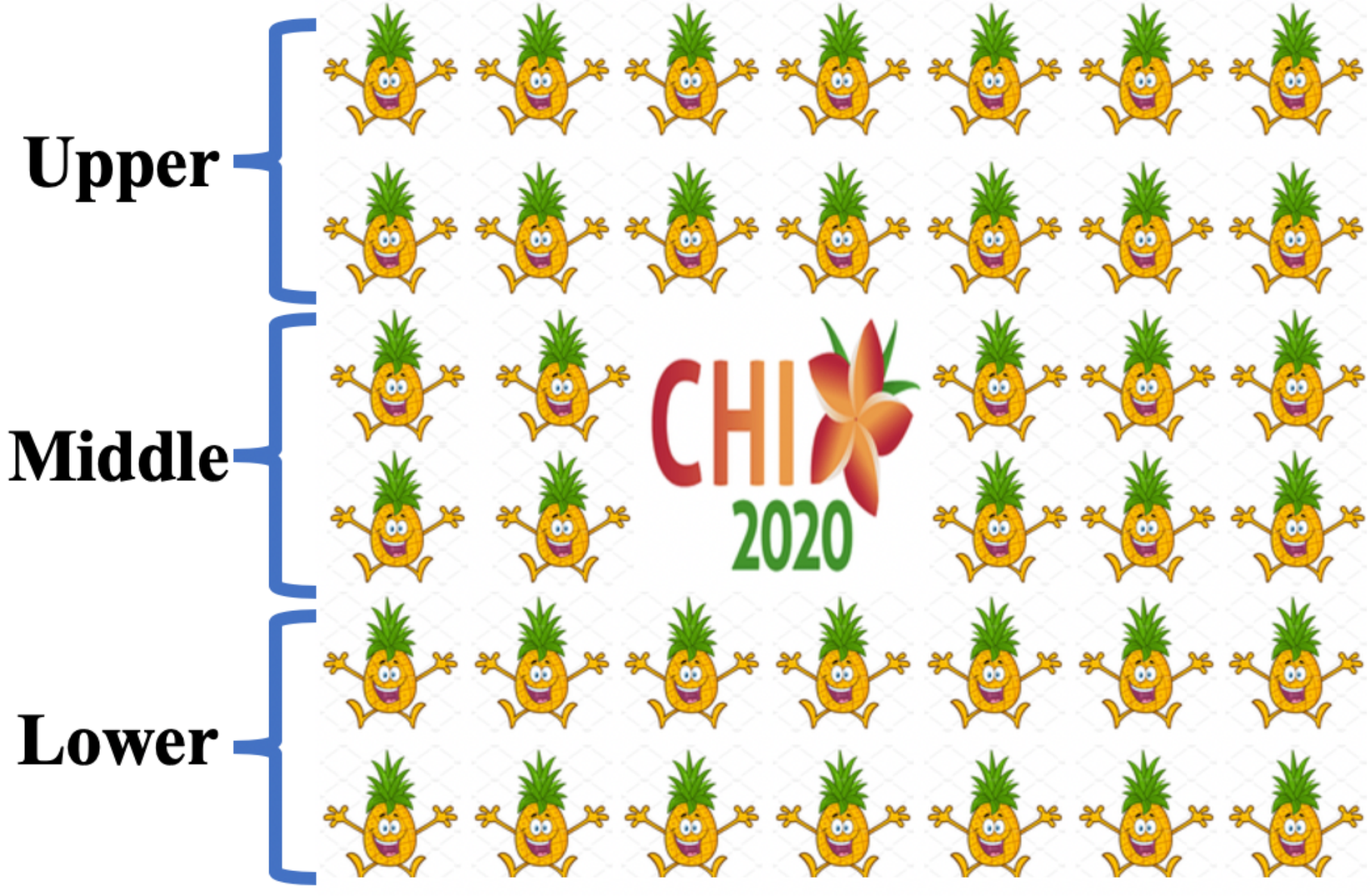}
    \end{minipage}
       &
    \begin{minipage}{0.345\textwidth}
{
   \begin{enumerate}
   \setlength{\itemsep}{0pt}
  \setlength{\parskip}{0pt}
	\item Divide the sub-layout into three areas: Upper, Middle, and Lower.
	\item Process the three areas as three connected flows using the approach described above.
\end{enumerate}
}
    \end{minipage}
    &
\begin{minipage}{0.345\textwidth}
      \includegraphics[width=63mm]{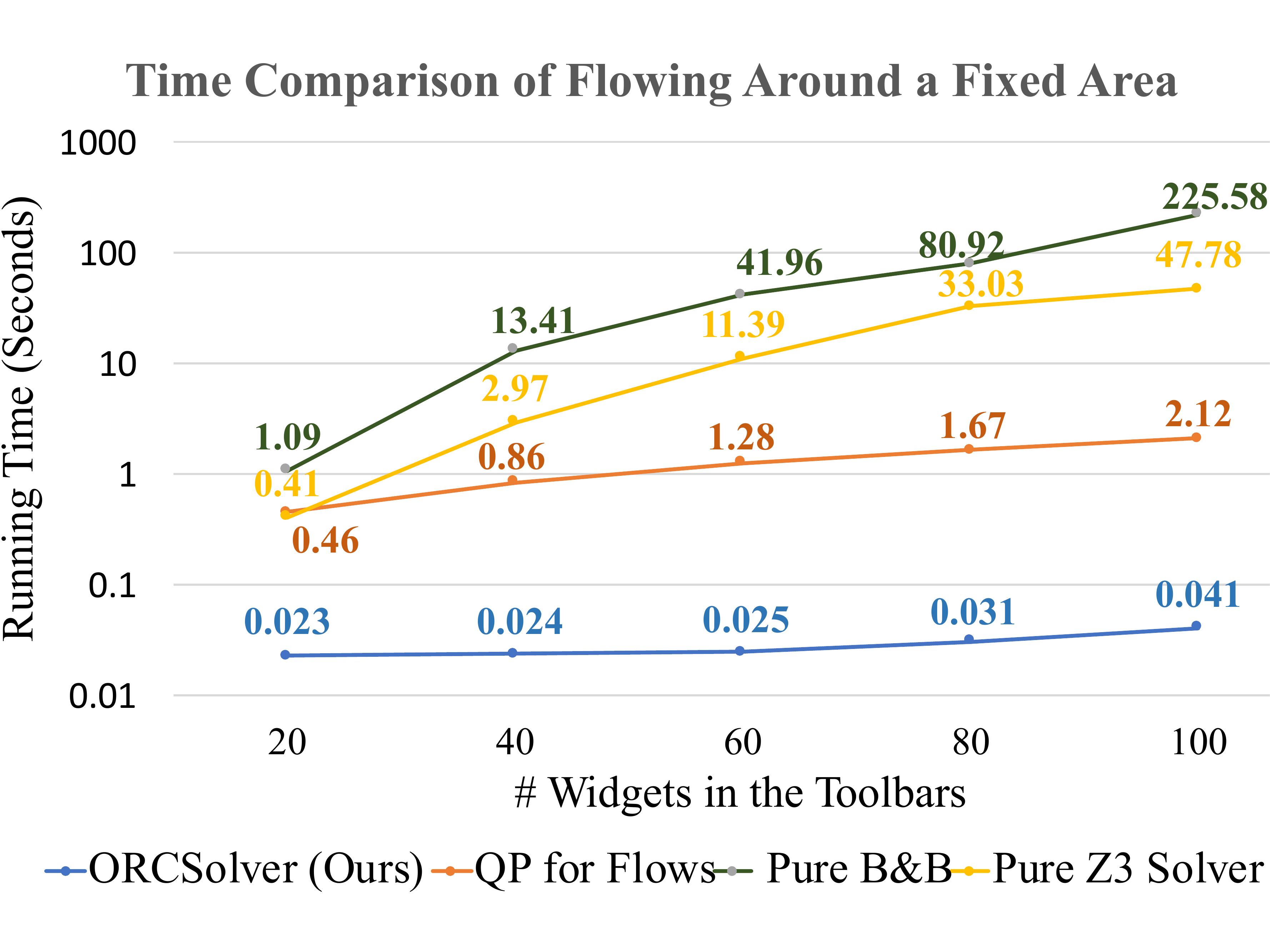}
    \end{minipage}
  
    \\ \hline

  \end{tabular}
  \caption{Comparison between ORCSolver and other approaches for simple, connected flow, and flowing around a fixed area.}\label{tbl:compare1}
\end{table*}

\section{Optimizing Flow Layouts}

{ORC layout unifies constraint-based layout and flow layout \cite{jiang2019ORC}. Optimisation of flow layouts is a central part of ORCSolver that addresses many ORC layout patterns.} In the following we describe the heuristic algorithm ORCSolver uses to locally optimise sub-layouts involving flow in linear time, in order to reduce their constraints. Without loss of generality and to keep the description simple, we discuss only horizontal flows as vertical flows are analogous. Furthermore, although we call the elements arranged in flows `widgets' as this is the most common use case, they could be other sub-layouts, which means the approach might be applied recursively.

Given an estimate of width based on firm edges, ORCSolver first uses heuristics to compute all possible numbers of rows in a horizontal flow, and then, given a specific number of rows, ORCSolver computes the optimal arrangement of widgets in the flow.
ORCSolver can estimate minimum, preferred, and maximum numbers of rows for a flow layout quickly by considering all contained widgets in their minimum, preferred, and maximum widths, respectively. This range of possible numbers of rows is then explored in the B\&B process by repeatedly applying \autoref{alg:flow} with different $numRows$ values. Using a simple greedy heuristic, ORCSolver tiles one widget after another until it hits a firm edge, and then move on to the next row. To fill a row without leaving gaps, ORCSolver redistributes the sizes of widgets evenly over the available width to create a balanced appearance \cite{zeidler2012constraint}.


\begin{table*}[h!]
  \centering
 \begin{tabular}{ | p{0.245\textwidth} | p{0.345\textwidth} | p{0.345\textwidth} |}
    \hline
    \multicolumn{1}{|c|}{Layout Pattern}  & \multicolumn{1}{|c|}{Heuristics} & \multicolumn{1}{|c|}{Time Comparison} \\ \hline
    \begin{minipage}{0.245\textwidth}
    \centerline{{\bf Balanced Flow}}
    \vspace{1mm}
      \includegraphics[width=45mm]{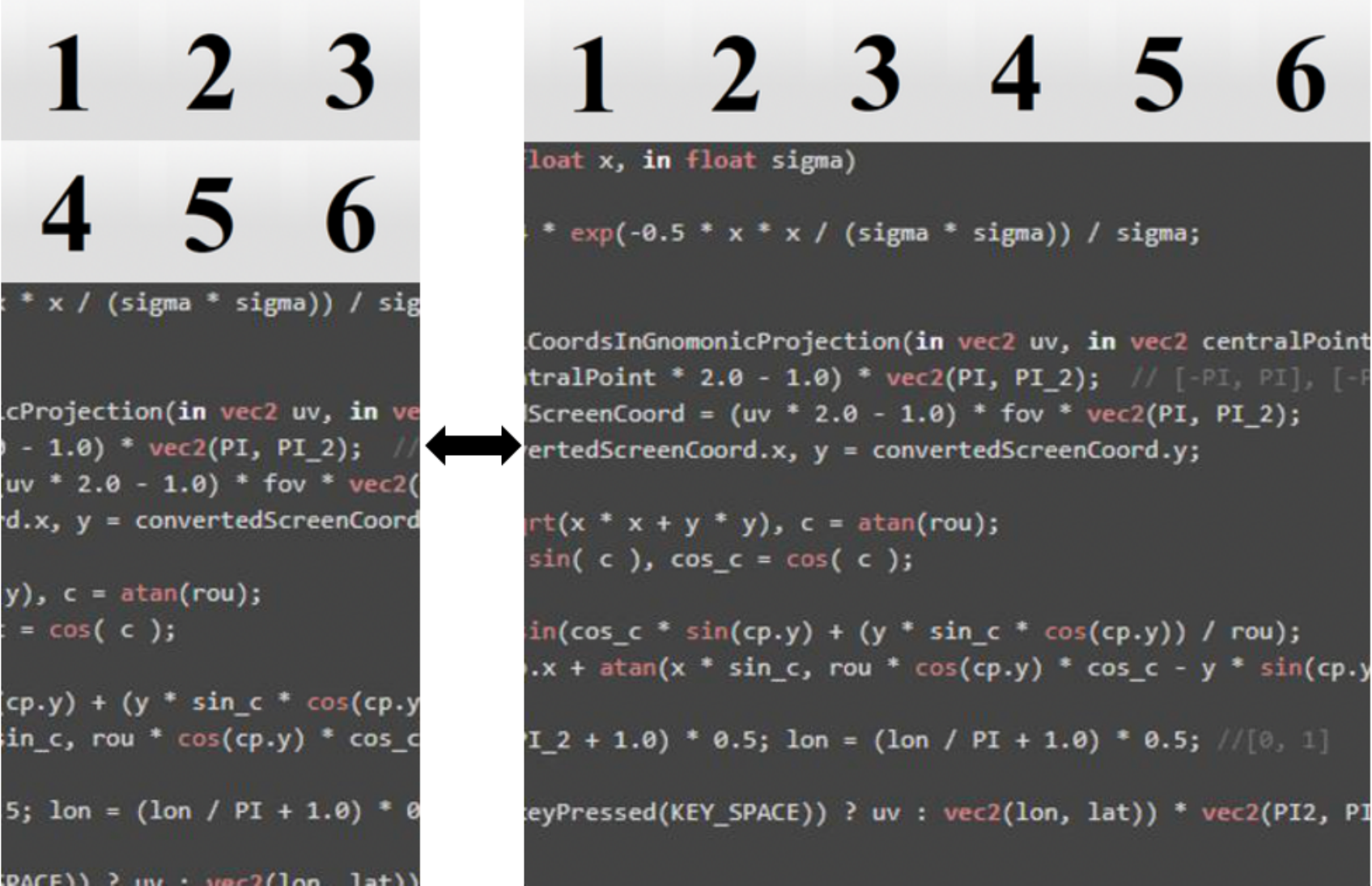}
    \end{minipage}
    &
    \begin{minipage}{0.345\textwidth}

{
All rows contain the same number of widgets.
   \begin{enumerate}
   \setlength{\itemsep}{0pt}
  \setlength{\parskip}{0pt}
	\item Compute all the factors of the total number of widgets in the flow sub-layout.
	\item Explore the different factors through B\&B, using appropriate numRows parameter values in the flow algorithm.
\end{enumerate}
}

    \end{minipage}
    &
\begin{minipage}{0.345\textwidth}
      \includegraphics[width=63mm]{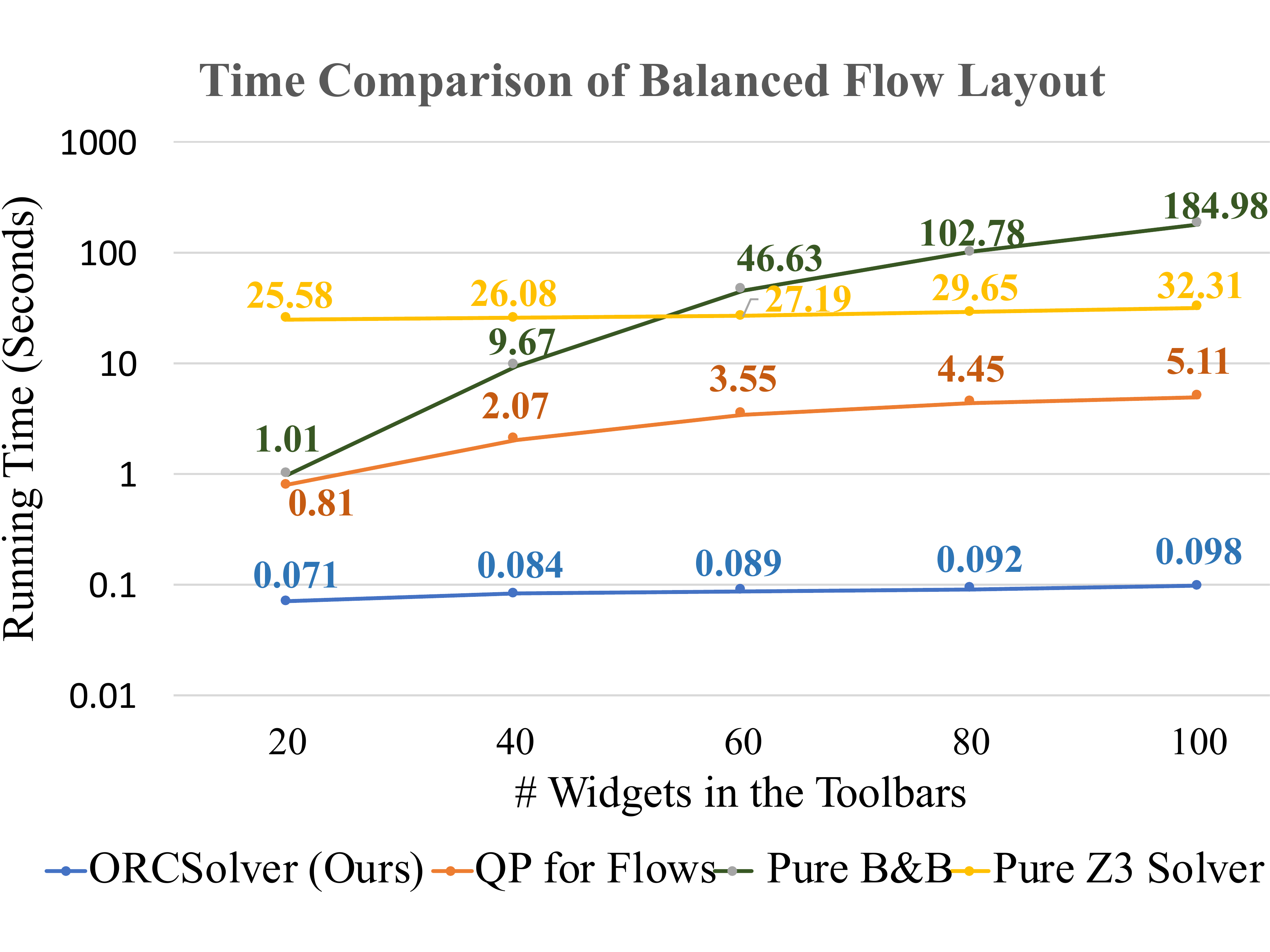}
    \end{minipage}
  
    \\ \hline
    
    \begin{minipage}{0.245\textwidth}
    \centerline{{\bf Optional Widgets}}
    \vspace{1mm}
      \includegraphics[width=45mm]{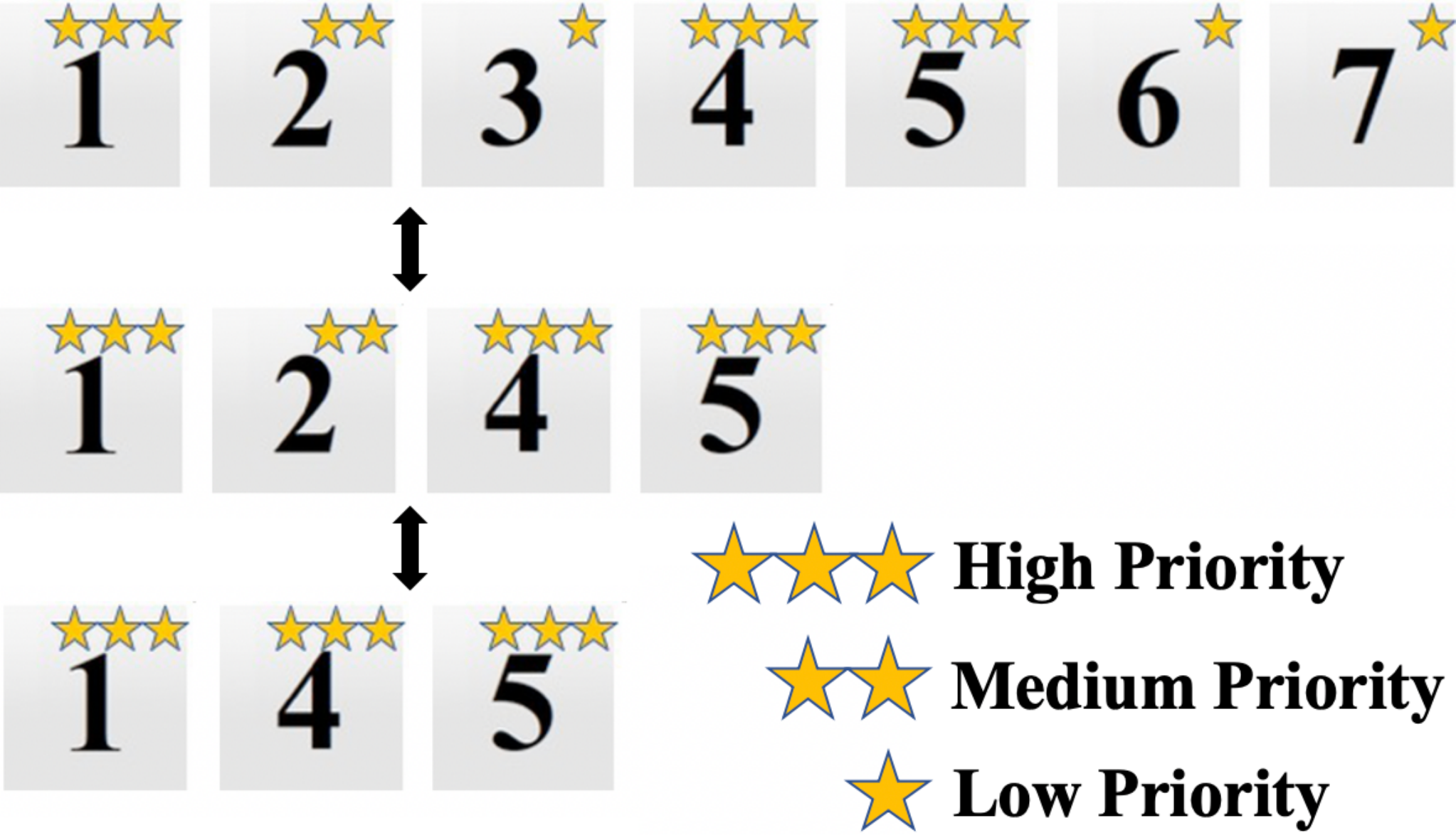}
    \end{minipage}
    &
    \begin{minipage}{0.345\textwidth}
    \vspace{1mm}
    {
    Widgets in flows can be optional. 
    Approach similar to Danzig's greedy algorithm \cite{danzig}:
   \begin{enumerate}
   \setlength{\itemsep}{0pt}
  \setlength{\parskip}{0pt}
	\item Sort all optional widgets by ascending weights and process flow row by row.
	\item If the total sum of widths of the widgets remaining to flow is greater than the available empty space in the flow (the estimated deviation $\Delta$ is negative) ,{\em i.e.}, not enough space in the remaining rows for all widgets, try removing the optional widget with the lowest priority, and then continue to remove widgets in ascending order of priority, as long as it reduces the loss.
	\item If $\Delta$ is positive ,{\em i.e.}, there is more available space for the remaining rows. check if we have removed optional widgets which can be added back in. We keep adding back optional widgets with the highest priority as long as it does not increase the loss.
\end{enumerate}
}


\vspace{1mm}
    \end{minipage}
    &
\begin{minipage}{0.345\textwidth}
      \includegraphics[width=63mm]{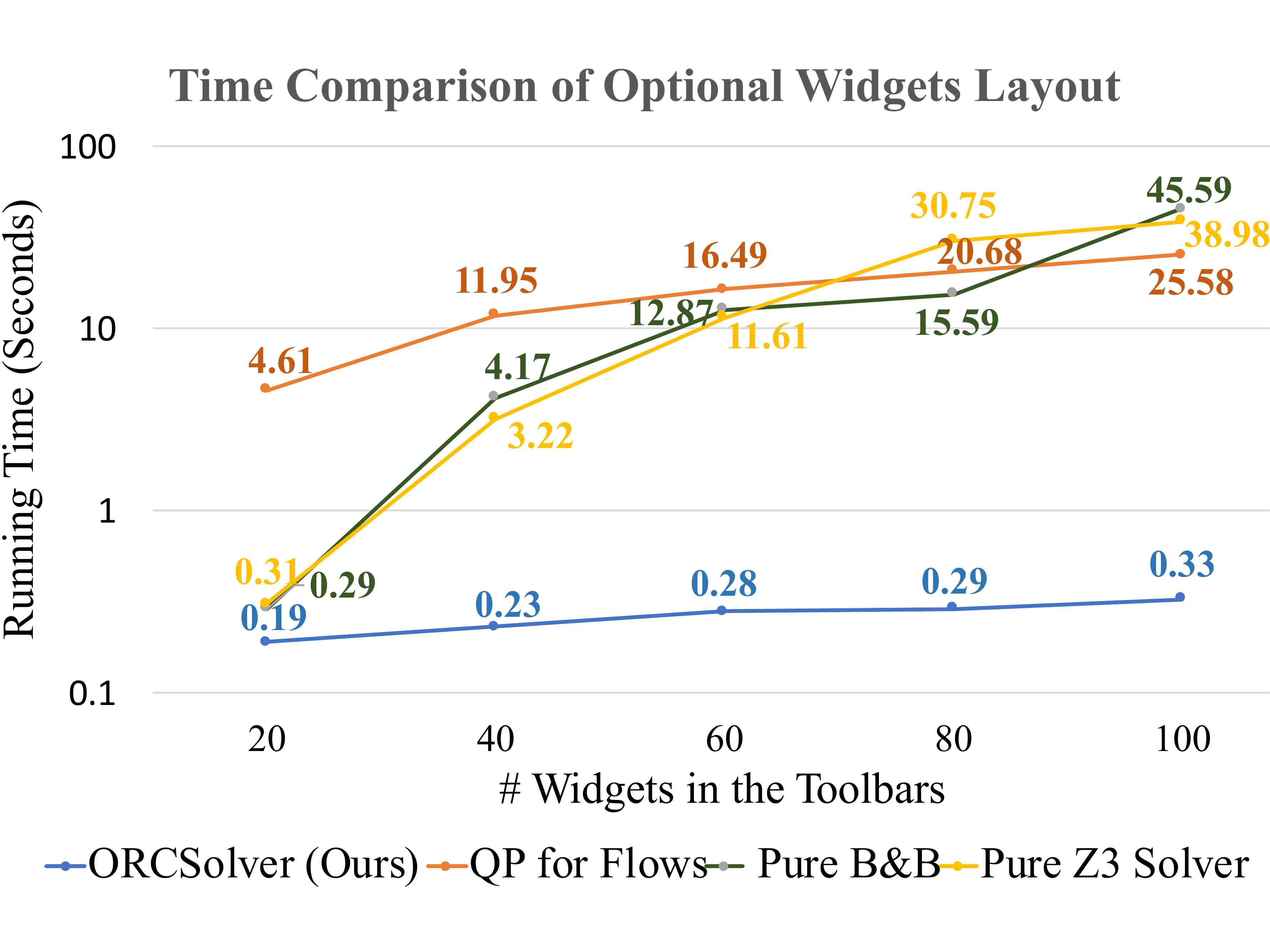}
    \end{minipage}
  
    \\ \hline
    
        \begin{minipage}{0.245\textwidth}
        \centerline{{\bf Alternative Positions}}
        \vspace{1mm}
      \includegraphics[width=45mm]{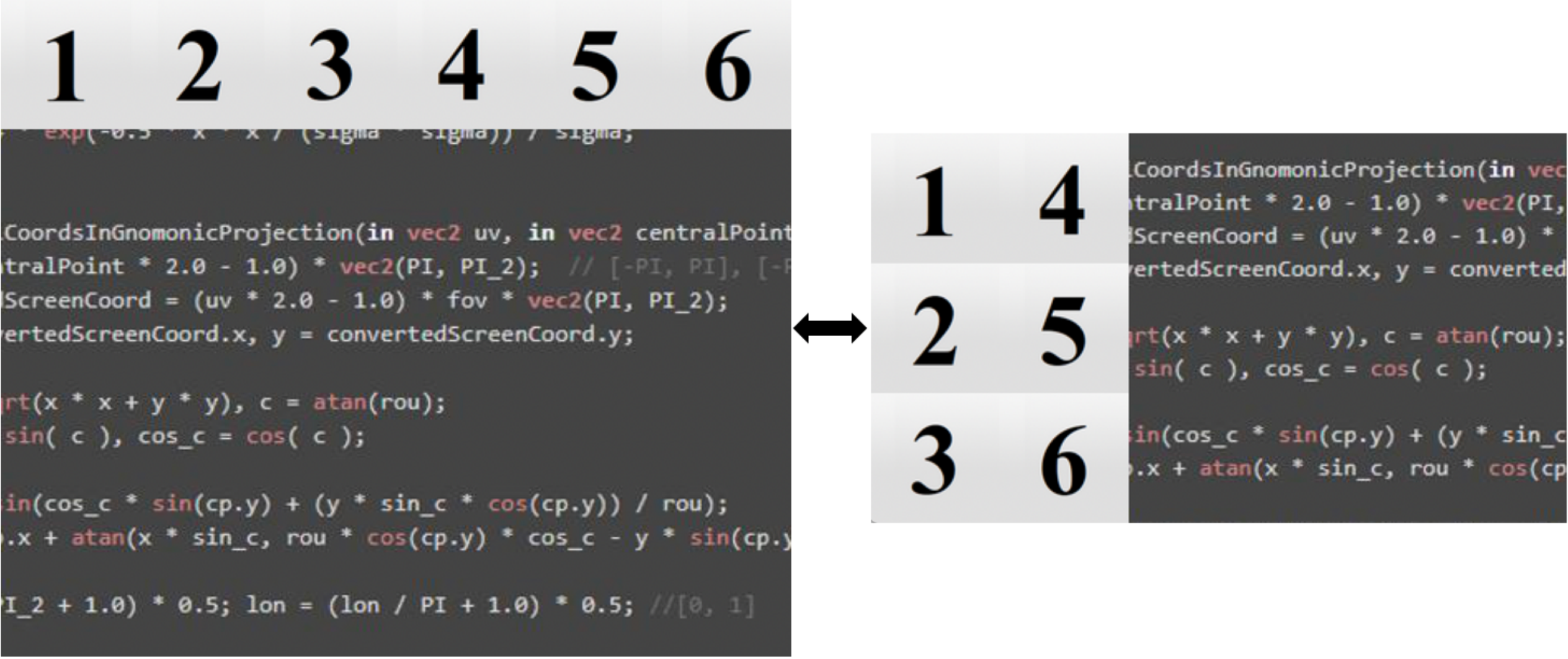}
     \end{minipage}
    &
    \begin{minipage}{0.345\textwidth}
    {
    Alternative positions can be defined for widgets or sub-layouts. For example, a top toolbar can move to the left of the window or vice versa when the screen size changes. The preferred alternative is given a higher weight. \\
    The final position is determined by B\&B minimizing the objective function.
    }
    \end{minipage}
    &
\begin{minipage}{0.345\textwidth}
      \includegraphics[width=63mm]{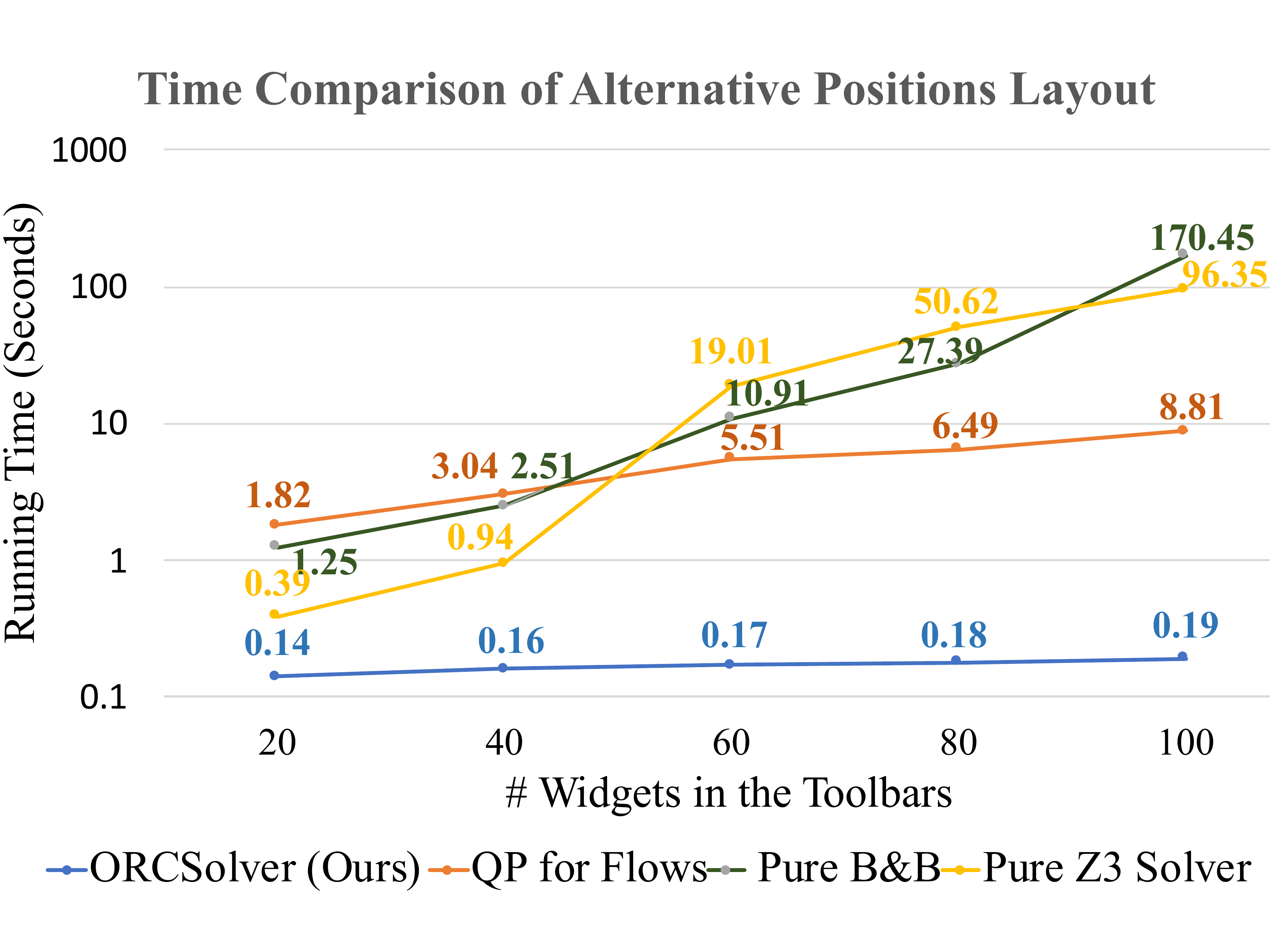}
    \end{minipage}
  
    \\ \hline

  \end{tabular}
  \caption{Comparison between ORCSolver and other approaches for balanced flow, optional widgets layout, and alternative positions patterns.}\label{tbl:compare2}
\end{table*}

\subsection{Objective Function}

To create a `balanced' layout appearance \cite{zeidler2012constraint},
our objective function for flows is the sum of the $L^2$ loss of the deviations between preferred and resulting sizes, resulting in a least-squares approach that reduces deviations by minimising their squares. Further, deviations from the preferred widget sizes can be scaled by widget weights:
\begin{dmath}
loss = \sum_i weight_i \times \big((width_i - prefWidth_i)^2 + (height_i - prefHeight_i)^2\big)
\end{dmath}

Instead of considering min/max constraints as hard constraints, we consider them as soft ones with very large weights. We add terms for them to the loss function, in a manner similar to the squared deviations from preferred sizes. This allows widgets to go beyond min/max sizes to avoid gaps in rows or prevent infeasible layouts, but incurs high penalties to make sure this is only done if there is no other solution. 
For the optional widget pattern, described below, we assigned a loss based on the omitted widgets' weights.
This ensures that the layout contains as many of those widgets as possible.

\begin{figure}[h!]
\centering
\includegraphics[width=\columnwidth]{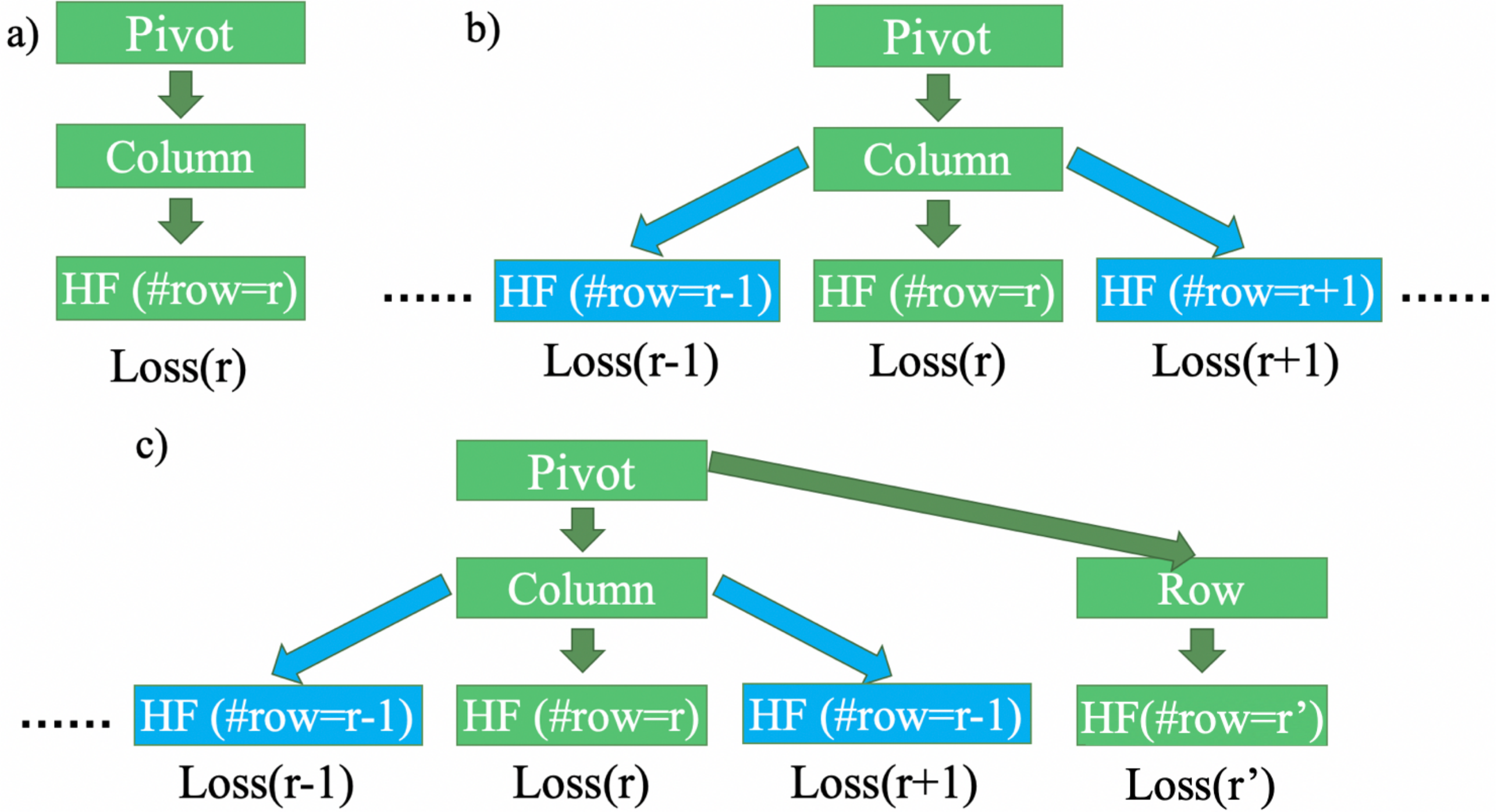}
\caption{Solving process for \autoref{fig:api_ex2}'s layout. The \textit{Pivot} layout has two alternatives, which ORCSolver processes individually to find the best solution.}
\label{fig:bb_ex2}
\end{figure}

\begin{figure*}[t]
\centering
\includegraphics[width=0.8\textwidth]{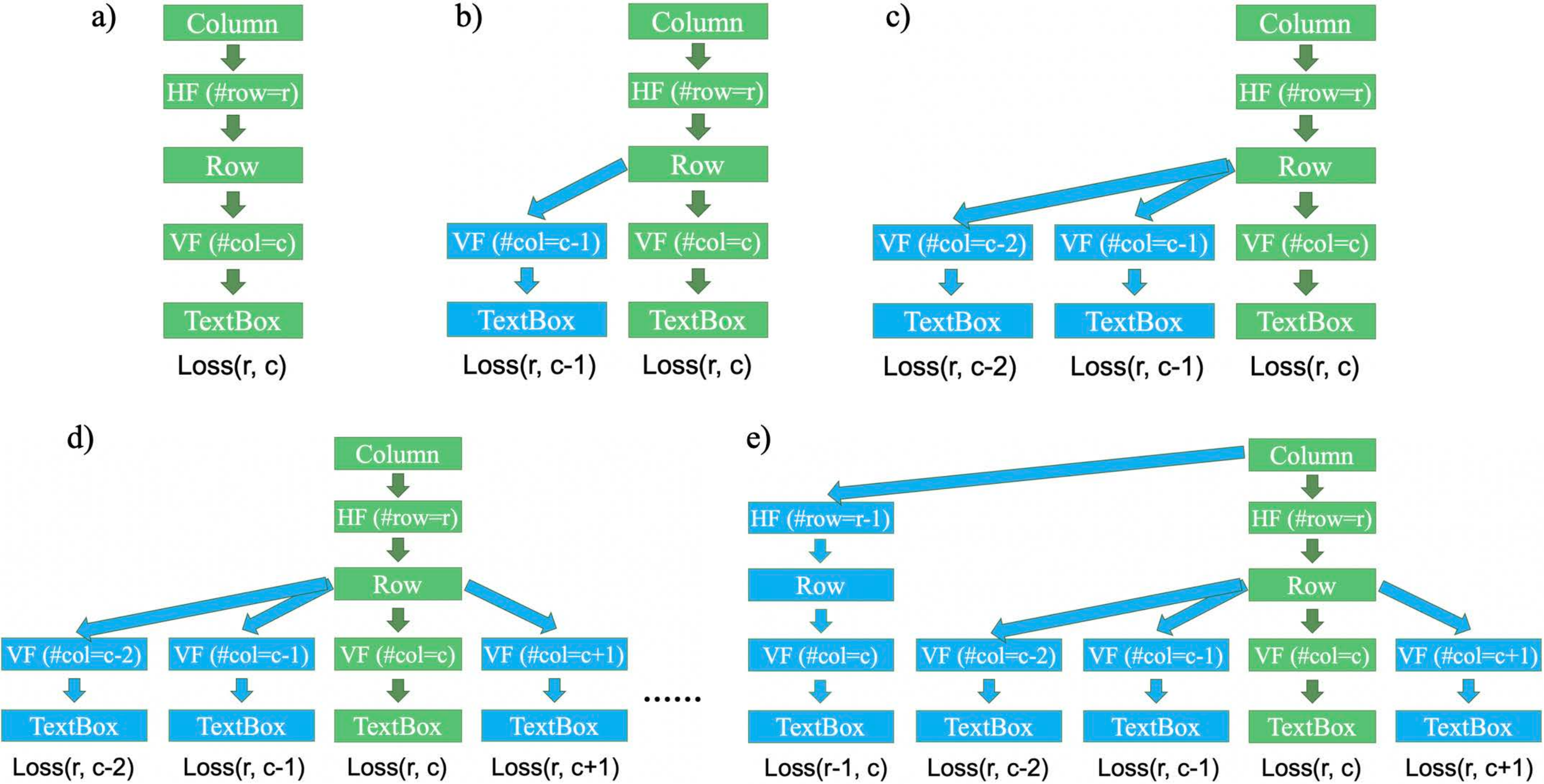}
\caption{The solving process for the layout shown in \autoref{fig:api_ex1}. After applying heuristics for flows, ORCSolver gets the preferred number of rows $r$ in the horizontal flow and the preferred number of columns $c$ in the vertical flow. a) ORCSolver constructs a tree for the layout based on the ORC notation and process it to get the penalty value for the main branch $Loss(r,c)$. b) ORCSolver adds a new branch for the vertical flow with $c-1$ columns and get its penalty value $Loss(r, c-1)$. If $Loss(r, c-1) \geq Loss(r, c)$, then it stops searching for smaller numbers of columns in the feasible range since the loss is already growing and smaller numbers can only lead to worse solutions. If $Loss(r, c-1) < Loss(r, c)$, ORCSolver stores this new branch as the currently best solution and c) move on to process smaller numbers of columns and repeat the same process. It stops trying smaller number of columns until a new branch is worse than its previous branch. d) ORCSolver tries larger numbers of columns in the same way. After finishing processing the vertical flow sub-layout, e) ORCSolver backtracks to the horizontal flow sub-layout and repeat the same process.}
\label{fig:bb_ex1}
\end{figure*}

\section{Global Optimisation Using Branch \& Bound}

ORCSolver constructs the B\&B tree for a layout based on its formal ORC layout notation. The tree is initially constructed as a one-branch tree, containing sub-layouts in the order in which they are chosen for processing based on their firm edges. For instance, the tree for the layout in \autoref{fig:api_ex1} with the notation {\it Column(HorizontalFlow, Row(VerticalFlow, TextBox))} can be represented as shown in  \autoref{fig:bb_ex1} a). Similarly, the tree for the layout in \autoref{fig:api_ex2} with the formalization {\it Pivot(Column(HorizontalFlow, TextBox))} can be represented as shown in \autoref{fig:bb_ex2} a).

Starting from the root node, as ORCSolver goes down the tree, it adds more and more constraints to the linear system. Each node in the B\&B tree has a layout specification containing variables, constraints and objectives. As ORCSolver goes down the tree, it clones the specification from the parent node and add simplified variables, constraints, and objectives for the current processed sub-layout based on the preprocessing results ({\em e.g.,} the heuristics for flows). ORCSolver solves the specification using a quadratic solver in the leaf node to calculate its overall loss value. If a node still contains OR-constraints after simplification, the Z3 solver is used for the solving step. Whenever ORCSolver reaches a node with a larger penalty value than the currently best leaf, it removes the entire sub-tree rooted at that node since the loss increases further as it goes down the tree, with more sub-layouts being processed to contribute to that loss. When hitting a leaf node, ORCSolver compares its loss with the currently best solution. If smaller, then the leaf is stored as the new best solution.

\autoref{fig:bb_ex2} shows the B\&B tree when ORCSolver solves the layout in \autoref{fig:api_ex2}, {\em i.e.}, \textit{Pivot(Column(HorizontalFlow, TextBox))}. The root contains the original layout formalization. The \textit{Pivot} layout contains two alternatives, {\em i.e.}, \textit{toolbar at the top} or \textit{toolbar on the left}. The solver uses B\&B to process the first alternative, starting with the \textit{toolbar at the top}, which means the solver adds a child node to the root node containing a simplified specification, {\em i.e.}, \textit{Column(HorizontalFlow, TextBox)}. The solver then moves on to the next level of the tree, the {\it HorizontalFlow}, which contains a set of toolbar widgets. The solver applies the heuristic solving module for the {\it HorizontalFlow} and adds a new child node with a simplified specification, {\em i.e.}, \textit{Column(heuristics result for HorizontalFlow, TextBox)}. The solver also computes the loss of this node according to the objective function. Then the solver moves on to process the next level for the {\it TextBox}, which is a leaf node of the solving tree. The layout does not contain any OR-constraints at this node, so ORCSolver can use a normal quadratic programming solver to calculate the final loss value for the layout. Once it reaches a leaf node, a global variable is used to maintain the currently best solution with the smallest loss value for the whole layout. Whenever the loss of a reached node is larger than the penalty of the currently best solution, ORCSolver ignores the entire subtree of that node. Then it backtracks up to the \textit{Pivot} node and process the other alternative, {\em i.e.}, \textit{toolbar on the left}. ORCSolver processes this branch as before. Whenever it reaches the bottom of the tree, ORCSolver checks whether it has smaller penalty value than the currently best solution and if so, overwrite the best solution.

\subsection{Discrete Gradient Search}

As part of the B\&B process, the solver conducts a gradient search for better solutions based on parameters of the heuristics modules, such as the number of rows or columns in flows. For example, as shown in \autoref{fig:bb_ex1}, when ORCSolver solves the layout in \autoref{fig:api_ex1}, the solver first processes the horizontal flow sub-layout since it is the first sub-layout in the \textit{Column} layout. It starts with the preferred number of rows
(\textit{numRows$_{pref}$}) to process the horizontal flow and get a good local fit. Then it moves on to the next sub-layout in the \textit{Column} which is the \textit{Row}. ORCSolver processes the \textit{Row} from left to right, so the next sub-layout to process is the vertical flow. Similarly, it starts with the preferred number of columns (\textit{numCols$_{pref}$}). Finally, it adds the {\it TextBox} to the window. If the remaining space is smaller than the {\it TextBox}'s minimum size, there is no feasible solution. As ORCSolver has reached a leaf of the tree, if a feasible solution exists, the solver minimizes the quadratic objective function globally and stores the result as the currently best solution. Then the solver backtracks in the B\&B tree to the {\it VerticalFlow} node and tries other alternatives that may lead to a better solution. It starts with the alternatives with fewer columns, {\em i.e.}, in the order of \textit{numCols$_{pref}$ - 1}, \textit{numCols$_{pref}$ - 2}, etc. We re-process the {\it VerticalFlow} with \textit{numCols$_{pref}$ - 1} columns first, and go down the tree to the {\it TextBox}. If we did not get a feasible solution in the previous branch, there may be a feasible solution in this branch, since, as we squashed the vertical flow area, there is now more space for the {\it TextBox}. Then, our currently best solution would be the leaf of this branch. If it was feasible in the previous branch, then the overall objective value may be smaller now. If it is smaller, we update the currently best solution accordingly and process the next {\it VerticalFlow} alternative with \textit{numCols$_{pref}$ - 2} columns. If that is not better than the the previous branch, ORCSolver stops trying even fewer columns since that would continue the trend of a narrower, taller {\it VerticalFlow} and lead to worse results. 
Analogously, it also tries increasing numbers of columns starting from \textit{numCols$_{pref}$+1} until that also increases the loss. 

The overall idea of our approach is to first try the local best fit with the preferred number of rows/columns in the flow at each flow node and then gradually explore into the direction of decreasing overall loss by adding new branches to process for both smaller and larger numbers of rows/columns. After getting the best solution for the {\it VerticalFlow}, ORCSolver backtracks to process the {\it HorizontalFlow}. Similarly, our solver gradually explores in the directions of decreasing overall loss values. For each branch created for the {\it HorizontalFlow}, it proceeds down the branch following the same process for the {\it VerticalFlow} as in the main stem. By backtracking and exploring possible better solutions, it optimises the solution for the overall layout.

\subsection{Constraint Solving}

Each node in the B\&B tree inherits all the variables, constraints and objectives from its parent and creates new boundary variables (top, bottom, left, right) for the corresponding sub-layout. If a sub-layout $S$ has min/pref/max sizes, we attach the following constraints to it: 
\begin{equation*}
 \begin{split}
&S.right - S.\mathit{left} = S.\mathit{prefWidth}\  [SOFT]\\
&S.bottom- S.\mathit{top} = S.\mathit{prefHeight}\ [SOFT]\\
&S.right - S.\mathit{left} \geq S.minWidth\\
&S.bottom - S.\mathit{top} \geq S.minHeight\\
&S.right - S.\mathit{left} \leq S.maxWidth\\
&S.bottom - S.\mathit{top} \leq S.maxHeight
\end{split}
\end{equation*}
As a quadratic programming solver cannot handle soft constraints directly, and instead of adding the soft constraints to the system, we rewrite the two soft constraints by adding slack variables $\delta_1$ and $\delta_2$: 
 \begin{equation*}
 \begin{split}
&S.right - S.\mathit{left} + \delta_1 = S.\mathit{pref\_width}  \\
&S.bottom- S.\mathit{top} + \delta_2 = S.\mathit{pref\_height} 
\end{split}
\end{equation*}
Similarly, we add slack variables to all other soft constraints. In addition, we add squares of all slack variables into the objective function, {\em i.e.},
with N soft constraints with slack variables $\delta_1$, $\delta_2$, $...$, $\delta_N$, we have the following objective function:
 \begin{equation*}
Minimize\ \ \delta_1^2 + \delta_2^2 + ... + \delta_N^2
\end{equation*}
ORCSolver runs the quadratic programming solver to solve the constraint system based on the variables, constraints, and objectives in a leaf node and get an objective value as the loss for the node to drive the B\&B process.

%% file: 05-implementation.tex
\section{Implementation}

We implemented our solver in \textit{Python} 
using a state-of-the-art quadratic programming solver, \textit{OSQP} \cite{osqp}, as our default solver.
Solving with B\&B is performed recursively with a method $solve()$ that is implemented by all layout types, which makes it fairly easy to extend ORCSolver with new layout patterns and heuristics. {ORCSolver with its API is available as open source from \url{https://github.com/YueJiang-nj/ORCSolver-CHI2020}.}




%% file: 06-evaluation.tex
\section{Evaluation}

To compare ORCSolver and three other approaches to solve layout specifications with OR-constraints, we measured the runtimes required to solve ORC layouts. We conducted the experiments on a laptop with an Intel i5 CPU and measured the average execution time over 10 runs {each with random feasible device sizes and random number of widgets in each flow, while keeping the total number of widgets constant. We found that the time varied only little between device sizes and was mainly determined by the total number of widgets}.
To avoid users touching complex constraints directly, Jiang et al.\ \cite{jiang2019ORC} proposed to describe layouts with ORC layout patterns. We based our work on the same approach and evaluated ORCSolver on six widely used layout patterns, including simple flow, connected flow, flow around a fixed area  (all in \autoref{tbl:compare1}), balanced flow, optional widgets, and alternative positions (all in \autoref{tbl:compare2}). {Most of these layouts are widely used in document layout and are not feasible with previous GUI layout systems.}


The work on ORC layouts \cite{jiang2019ORC} used Z3 \cite{de2008z3} to solve layout specifications with OR-constraints. Although various Mixed Integer Programming (MIP) and Satisfiability Modulo Theories (SMT) solvers exist for constraint-based systems, Z3 is the only solver that can deal with disjoint disjunctions and solve constraint systems with OR-constraints efficiently. {Z3 dominates the competition in the SMT-COMP events \cite{SMT2017}
, as it is typically much faster than most competitors on most tasks.} We compare ORCsolver with the following three approaches:
\begin{description}
\item[Pure Z3 Solver] This is the original solving technique for ORC layouts \cite{jiang2019ORC} using the state-of-the-art Z3 solver for OR-constraint systems.

\item[QP for Flows] Each widget in flows has min/pref/max size constraints. {This approach solves all these constraints in B\&B nodes using the state-of-the-art OSQP quadratic programming solver, after applying {our} heuristics simplifying the flow constraints.} 

\item[Pure B\&B] In addition to using B\&B for optimising layout parameters, such as the number of rows/columns in flows, this approach also uses B\&B to optimise the internal structure of the layouts, {\em e.g.}, solves flows with each branch representing different numbers of widgets in each row/column. 
\end{description}

Our API allows us to plug in different solvers for different ORC patterns (including ORCSolver, ``QP for Flows'', and ``Pure B\&B'').
As shown in the logarithmic plots in \autoref{tbl:compare1} and \autoref{tbl:compare2}, {ORCSolver can solve layout patterns at near-interactive rates and about two orders of magnitude faster than other methods, especially when the number of widgets is large.}

{\subsection{Differences in Outcomes}}

\begin{figure}[t]
\centering
\includegraphics[width=0.9\columnwidth]{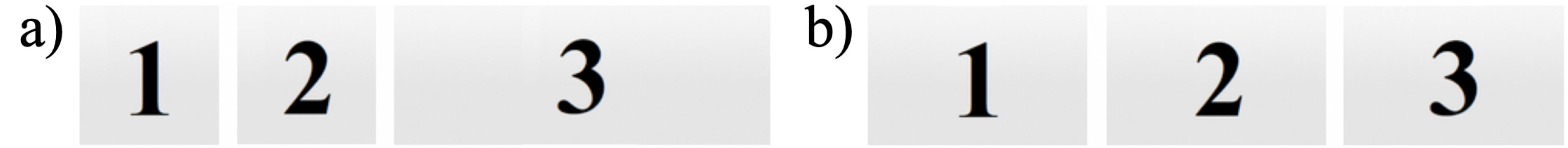}
\caption{Layout results with a) Z3 Solver and b) our ORCSolver.}
\label{fig:comparison}
\end{figure}

\begin{figure}[t]
\centering
\includegraphics[width=0.7\columnwidth]{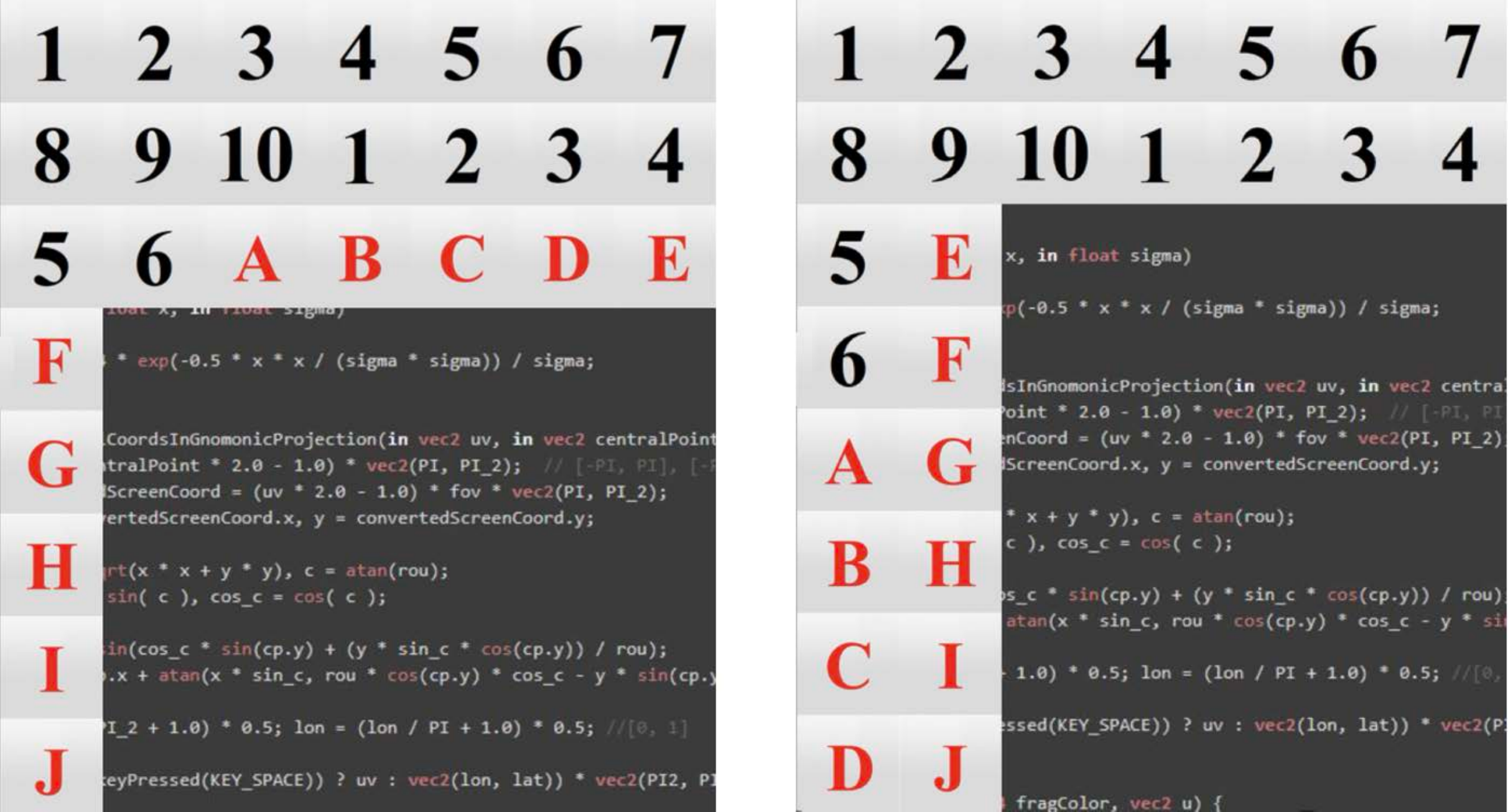}
\caption{Different possible outcomes may have the same objective value.}
\label{fig:diff_outcomes}
\end{figure}

{Different solving approaches may lead to different resulting layouts. Z3's default \textit{WMax} \cite{nieuwenhuis2006wmax} optimization, which was used in the original ORC layout work \cite{jiang2019ORC},
is fast but does not balance deviations between violated constraints, which leads to an unbalanced, less aesthetic appearance \cite{zeidler2012constraint}. For example, as shown in \autoref{fig:comparison}, if three widgets in a row have the same preferred size and the row width is not enough or too large for all of them to have their preferred size, then \textit{WMax} \cite{nieuwenhuis2006wmax} tries to make two of them have the preferred size to satisfy as many constraints as possible, which may lead to a bad layout result (\autoref{fig:comparison}a). One can avoid this problem by applying a quadratic objective function within Z3, but this is prohibitively slow (several minutes for 10+ widgets) and none of Z3's other optimizers improves this. Many differences of Z3's outputs are due to its inability to solve quadratic functions tractably. By using quadratic programming, our solver distributes the deviations over the widgets and balances their appearance (\autoref{fig:comparison}b). Thus, ORCSolver is not only faster than Z3, but also able to produce more balanced and aesthetic results. As complex layouts are often underspecified, several possible optimal solutions may exist. For example, as shown in \autoref{fig:diff_outcomes}, placing the buttons either in the third row or a second column may have the same objective value, and different solvers may make different choices.
The general challenge of underspecification can be addressed with suitable GUI editors \cite{zeidler2013auckland, zeidler2017tiling}. }

\subsection{Computational Complexity}
 
In the B\&B process we start with a one-branch search tree and add more branches for different sub-layout parameters such as flows with different numbers of rows/columns. Yet, the number of branches we add is often very small as it is bounded by the feasible parameter range, {\it e.g.}, the minimum to maximum numbers of rows/columns in flows, and we stop adding branches in either direction (smaller or larger than the preferred number) once the previous branch has larger loss than the current best result. Hence, we only have an approximately constant factor of branches to add. The time complexity to process each branch from root to leaf is $O(numSublayouts)$. For each leaf, we use quadratic programming to solve the simplified layout specification, which has average polynomial complexity and is roughly linear in practice. Our heuristics for solving ORC patterns are $O(numWidgets)$. Therefore the overall complexity of ORCSolver is approximately linear in practice. By comparison, the original ORCLayout approach \cite{jiang2019ORC} used only Z3, which exhibits \textit{exponential} runtime. Since ``QP for Flows'' solves a quadratic programming problem for each B\&B node, its complexity is at least quadratic. ``Pure B\&B'' is a brute-force method with exponential complexity.

%% file: 07-conclusion.tex
\section{Discussion \& Conclusion}

In this paper, we presented ORCSolver, an efficient solver for adaptive GUI layout with OR-constraints. Our approach overcomes the main performance bottleneck for using OR-constrained layouts, and enables ORC systems to be solved efficiently at {near-interactive} rates even with large numbers of widgets. 
A formal notation for ORC layouts enables ORCSolver to support various types of layout patterns efficiently. Our modular approach of defining sub-layouts and corresponding solvers creates a framework that allows for further expansion, {\it e.g.}, for different layout patterns and solving strategies.

The novel ORCSolver approach is based on a branch\&bound algorithm using interval arithmetic on layout parameters to limit the branching factor, so that OR-constraint systems for GUI layouts can be solved at {near-interactive} speeds.
As illustrated through the performance graphs, our approach can solve various layout patterns much faster than previous work, typically by two orders of magnitude for larger layouts, which is a major improvement that becomes more important as GUI layouts get more sophisticated. The results show that it is now in principle feasible to solve ORC layouts on mobile devices.

Efficient solving of constraint-based layout systems also enables interactive modification and/or adaptation of layouts.
Given that solving times reach at most 0.1 - 0.3 seconds, this means that ORCSolver can recompute layouts at speeds that are high enough to support a window resize or interactive GUI editing.
Finally, our new approach enables designers to utilize the benefits of the unification of conventional constraint-based and flow layouts, which opens up new possibilities for GUI design. This is especially important for the definition of flexible GUI layouts that adapt seamlessly to screens with different sizes, orientations, or aspect ratios -- while still being based on only a single layout specification.

\subsection{Limitations and Future Work}

ORCSolver solves a layout based on the formal specification of said layout. Thus, users currently have to specify a layout manually before feeding it into ORCSolver. 
{Designers currently cannot visually specify ORC patterns through a GUI. Although there is a visual editor for ORC layouts, the editor does not yet allow users to specify resizing behaviour purely by direct manipulation. Inferring an ORC pattern from a layout is an avenue for future work.} 
{Additional heuristics such as meta-heuristics for optimization might further accelerate the solving process, but might also sacrifice accuracy. For example, large neighbourhood search \cite{pisinger2010large} could find good candidate solutions by prioritizing promising search paths.}
We could also provide more heuristic rules and let the user choose which heuristic they want to use, or choose heuristics automatically based on runtime constraints, which can make our framework more flexible and generic.
Finally, our current implementation does not support incremental solving. A warm-start solving option after a change in a layout is a useful functionality we plan to add in the future. {Adding an incremental approach might reach real-time performance.}